\documentclass[aps,twocolumn,prb,preprintnumbers, 10pt]{revtex4-2}
\usepackage[left=1.9cm, right=1.9cm, top=1.9cm, bottom=1.9cm]{geometry}
\usepackage{xpatch}
\makeatletter
\patchcmd{\@ssect@ltx}
    {\addcontentsline{toc}{#1}{\protect\numberline{}#8}}
    {}
    {}
    {}
\makeatother
\pdfoutput=1
\usepackage{color}
\usepackage{enumitem}
\usepackage[dvipsnames]{xcolor}
\usepackage{hyperref}
\definecolor{amaranth}{rgb}{0.9, 0.17, 0.31}
\definecolor{forestForestGreen(web)}{rgb}{0.13, 0.55, 0.13}
\definecolor{dgreen}{rgb}{0.13, 0.55, 0.13}
\definecolor{blue(munsell)}{HTML}{005567}
\definecolor{oxfordblue}{rgb}{0.0, 0.2, 0.4}
\definecolor{SymTFTColores}{rgb}{0, 0.33, 0.701}
\definecolor{bblue}{rgb}{0.0, 0.58, 0.71}
\hypersetup{
pdfstartview={FitH}, 
pdftitle={}, 
colorlinks=true, 
linkcolor=oxfordblue, 
citecolor=oxfordblue, 
filecolor=oxfordblue, 
urlcolor=oxfordblue
}
\usepackage{pgfplots}
\pgfplotsset{compat=1.18}
\usepackage{amssymb}
\usepackage{amsfonts}
\usepackage{graphicx}
\usepackage{epstopdf}
\usepackage{dcolumn}
\usepackage{amsmath}
\usepackage{latexsym,bm}
\usepackage{amsthm}
\usepackage{slashed}
\usepackage{float}
\usepackage{color}
\usepackage{url}
\usepackage{longtable}
\usepackage{rotating}
\usepackage[normalem]{ulem}
\usepackage{faktor}
\usepackage{tikz-cd}
\usepackage{tensor}
\usepackage{array}
\usepackage{tabularx}
\usepackage{makecell}
\usepackage{changepage}
\usetikzlibrary{fit}

\newcommand{\om}{\omega}

\usepackage{tikz}
\usepackage{diagbox}
\usetikzlibrary{positioning}
\usetikzlibrary{calc}
\usetikzlibrary{decorations.pathreplacing,calligraphy}

\usepackage{xstring}
\usetikzlibrary{decorations.pathmorphing} 
\usetikzlibrary{decorations.markings} 
\usetikzlibrary{arrows} 
\usetikzlibrary{shapes} 
\usetikzlibrary{matrix} 
\usetikzlibrary{positioning} 
\usepackage[english]{babel} 
\usepackage[autostyle]{csquotes}
\usepackage{pifont}
\usetikzlibrary{shapes.multipart}

\tikzset{->-/.style={decoration={
  markings,
  mark=at position .5 with {\arrow{>}}},postaction={decorate}}}

\newcommand{\bea}{\begin{eqnarray}}
\newcommand{\eea}{\end{eqnarray}}
\newcommand{\be}{\begin{equation}}
\newcommand{\ee}{\end{equation}}
\newcommand{\ba}{\begin{aligned}}
\newcommand{\ea}{\end{aligned}}
\newcommand{\bit}{\begin{itemize}}
\newcommand{\eit}{\end{itemize}}
\newcommand{\ben}{\begin{enumerate}}
\newcommand{\een}{\end{enumerate}}
\newcommand{\nn}{\nonumber}

\newcommand{\id}{\text{id}}

\newcommand{\GS}{\text{GS}}

\newcommand{\lb}{\left(}
\newcommand{\rb}{\right)}
\newcommand{\lbb}{\left[}
\newcommand{\rbb}{\right]}

\newcommand{\obj}{\text{obj}}

\newcommand{\half}{\frac{1}{2}}

\newcommand{\Z}{{\mathbb Z}}

\newcommand{\bC}{{\mathbb C}}
\newcommand{\Q}{{\mathbb Q}}

\newcommand{\cA}{\mathcal{A}}

\newcommand{\cC}{\mathcal{C}}

\newcommand{\cL}{\mathcal{L}}

\newcommand{\cO}{\mathcal{O}}
\newcommand{\cP}{\mathcal{P}}

\newcommand{\cS}{\mathcal{S}}

\newcommand{\cZ}{\mathcal{Z}}

\renewcommand{\ol}{\overline}

\newcommand{\Hom}{\text{Hom}}

\newcommand{\Rep}{\mathsf{Rep}}

\renewcommand{\dim}{\text{dim}}

\renewcommand{\Q}{\bm{Q}}
\newcommand{\sym}{\text{sym}}
\newcommand{\phys}{\text{phys}}

\newcommand{\GSr}{|{\GS^{(k)}\!,q}\rangle}

\newcommand{\GSkl}{\langle{\GS^{(k)}}|}
\newcommand{\GSkr}{|{\GS^{(k)}}\rangle}

\newcommand{\GSrz}{|{\GS^{(k)}\!,\id}\rangle}

\makeatother

\usepackage{physics}

\newcommand{\bbI}{\mathbb I}

\newcommand{\diag}{\text{diag}}

\newcommand{\II}{\text{II}}

\newcommand{\CFT}{\text{CFT}}

\usepackage{float}

\usepackage{makecell}
\makeatletter
\newcommand\xlabel[2][]{\phantomsection\def\@currentlabelname{#1}\label{#2}}
\makeatother

\newcommand{\SymCatS}{\cS}
\newcommand{\SymCatQ}{\Rep(Q_8)\rtimes {S_3}}

\makeatletter
\def\l@subsubsection#1#2{}
\makeatother

\begin{document}

\title{Twin Phases: 
Intrinsic Deconfined Quantum Criticality 
}

\author{Alison Warman}
\author{Yuhan Gai}
\author{Sakura Sch\"afer-Nameki}

\affiliation{\vspace*{0.5mm}
Mathematical Institute, University
of Oxford, Woodstock Road, Oxford, OX2 6GG, United Kingdom}


\begin{abstract} 
\noindent 
We introduce the concept of twin phases for a symmetry $\mathcal{S}$, defined as inequivalent phases, whose order parameters are part of the same generalized charge under $\mathcal{S}$. Stable, direct transitions between such twin phases are never spontaneous-symmetry-breaking transitions, even after (partially) gauging the initial symmetry $\mathcal{S}$: they are phase transitions without hidden symmetry breaking. We illustrate this with an (anomalous) finite group symmetry in 1+1d, which exhibits such intrinsically beyond Landau transition, i.e. an intrinsically Deconfined Quantum Critical Point (DQCP).  

\end{abstract}

\maketitle

\noindent\textbf{Introduction.}
The classical Landau paradigm \cite{Landau:1937obd} characterizes gapped phases of a finite group symmetry $G$ by their unbroken subgroup $H \leq G$, with  transitions between them requiring spontaneous symmetry breaking (SSB): if $H_1$ and $H_2$ are the preserved symmetries a continuous (second order) requires $H_1 < H_2$:
\be\label{Landau}
\begin{split}
  \begin{tikzpicture}
        \node[] (n1) at (0,0) {$G$};
        \node[] (n2) at (-1.00, -0.8) {$H_1$};
        \node[] (n3) at (1.00,-0.8 ) {$H_2$};
        \node at (0,-0.8) {$<$};
        \draw[->] (n1) -- (n2);
        \draw[->] (n1) -- (n3);
\end{tikzpicture}
\end{split}
\ee
In light of generalized symmetries and the extension to the Categorical Landau Paradigm \cite{Bhardwaj:2023fca}, it was 
recently conjectured that all 1+1d phase transitions are SSB transitions, possibly after gauging  
\cite{Chen:2025uno}. This is motivated by numerous examples \cite{Kennedy:1992ifl, kennedy1992hidden, Zhang:2022wwn, Chen:2025ivo}, and thus far even for phase transitions in the presence of categorical symmetries \cite{Schafer-Nameki:2023jdn, Shao:2023gho} has not been disproven \cite{Bhardwaj:2024qrf,  Bhardwaj:2024wlr, Bhardwaj:2024kvy, Chatterjee:2024ych, Warman:2024lir, Bottini:2025hri, Chen:2026fai}. 
In this paper we provide counterexamples: transitions with no hidden SSB 
description (there is no gauging to map them to Landau transitions), which are thus {\bf intrinsically beyond Landau 
transitions}. They correspond to deconfined quantum critical points (DQCPs) 
\cite{Senthil:2003eed, Senthil:2023vqd}, that remain DQCPs even after (partially) gauging the symmetry. We illustrate this phenomenon  with a finite group $G$.
The key insight comes from subtler aspects of gapped phases for finite group symmetries  \cite{Gai:2026hjk}: {mathematically they are described by} module categories, or in the Symmetry Topological Field Theory (SymTFT) \cite{Ji:2019jhk,  Gaiotto:2020iye, Apruzzi:2021nmk, Freed:2022qnc}, gapped boundary conditions (i.e. Lagrangian algebras). 
The main new concept is that of {\bf twin algebras} \cite{Gai:2026hjk} in the SymTFT, defined as algebras with the same anyon decomposition but different algebra structures \footnote{Equivalently, there  are twin module categories that are equivalent as linear categories but inequivalent as module categories over any symmetry related by gauging.}. Applied to the SymTFT classification of gapped phases suggests the following:

\vspace*{1mm}
\noindent{\bf Definition.}
\textit{\textbf{ Twin (gapped) phases} are two distinct (gapped) phases for a symmetry 
$\cS$ whose order parameters (OPs) lie in the same generalized charge under $\cS$.  \textbf{Twin phase transitions} are stable second-order phase transitions between twin phases.}\\
\noindent
Generalized charge refers to the multiplet structure under $\cS$ 
\cite{Bhardwaj:2023ayw}. Equivalently, twin phases arise from the same anyons in the SymTFT, but neverthelss have distinct OPs. 
If $\cS$ is a finite group $G$, the OPs of twin phases can e.g. be different 
components of a higher-dimensional irrep of $G$. 

The purpose of this paper is to illustrate the physical implications of twin algebras for theories with finite group symmetries (or gaugings thereof), and is thus firmly in the setting originally studied by Landau. 
Concretely we consider the group 
\be \label{GL23}
G= GL(2,3) \cong Q_8 \rtimes S_3 \,.
\ee
This group possesses twin gapped phases, but to obtain a stable, direct second-order phase transition, we  consider $G^\omega$ with anomaly $\omega \in H^3(G, U(1))$. 
This is a mixed anomaly between the preserved symmetries in the twin gapped phases, which we show leads to a stable second-order phase transition between them:
\be\label{TwinPhases}
\begin{split}
  \begin{tikzpicture}
        \node[] (n1) at (0,0) {$GL(2,3)^{\omega}$};
        \node[] (n2) at (-1.00, -1) {$S_3^{(1)}$};
        \node[] (n3) at (1.00,-1 ) {$S_3^{(2)}$};
        \node at (0,-1) {$\not<$};
        \node at (3.2, -0.5) {with $\quad  
        \ba 
        S_3^{(1)} &\cong S_3^{(2)}\cr 
        S_3^{(1)} &\not\sim S_3^{(2)}\ea $};
        \draw[->] (n1) -- (n2);
        \draw[->] (n1) -- (n3);
\end{tikzpicture}
\end{split}
\ee
Here $S_3^{(k)}$ are permutation groups on 3 elements,  each preserved in one of the two gapped phases, but they are {\bf non-conjugate} ($\not\sim$) subgroups in $G$. The OPs for each phase are different components of a higher-dimensional irrep (generalized charge) of $G$, resulting in two distinct gapped phases.

\begin{figure}
$$
\begin{tikzpicture}
\begin{scope}[shift={(0,0)}]
\draw [SymTFTColores, fill= SymTFTColores, opacity = 0.2] 
(0,0) -- (0,2) -- (2,2) -- (2,0) -- (0,0) ; 
\draw [white] (0,0) -- (0,2) -- (2,2) -- (2,0) -- (0,0)  ; 
\draw [very thick] (0,0) -- (0,2) ;
\draw [very thick] (2,0) -- (2,2) ;
\draw [ thick] (0,0.8) -- (2,0.8) ;
\fill[] (0,0.8) circle (0.05cm);
\fill[] (2,0.8) circle (0.05cm);
\node[above] at (1,0.8) {$a'$};
\draw [thick] (0,0.4) -- (2,0.4) ;
\fill[] (0,0.4) circle (0.05cm);
\fill[] (2,0.4) circle (0.05cm);
\node[above] at (1,0.4) {$a$};
\draw [thick]  (0,1.3) -- (2,1.3)  ;
\fill[] (0,1.3) circle (0.05cm);
\fill[] (2,1.3) circle (0.05cm);
\fill[] (2,1.3) circle (0.05cm);
\node[above] at (1,1.3) {$a''$};
\node[above] at  (1,2)  {$D^\omega(G)$} ;
\node[above] at (-0.2,2) {$\cL_\sym^{G,\om}$}; 
\node[above] at (2.2,2) {$\cL_\phys^{k=1,2}$}; 
\end{scope}
\begin{scope}[shift={(4,0)}]
\node at (-1,1) {$=$} ;
\draw [very thick] (0,0) -- (0,2) ;
\fill[] (0,1.3) circle (0.05cm);
\fill[] (0,0.8) circle (0.05cm);
\fill[] (0,0.4) circle (0.05cm);
\node[right]at  (0,1.3) {$\cO_{a''}^{(k)}$};
\node[right]at  (0,0.8) {$\cO_{a'}^{(k)}$};
\node[right]at  (0,0.4) {$\cO_a^{(k)}$};
\node[above] at (0.1,2) {$\cP_{k=1,2}$}; 
\end{scope}
\end{tikzpicture}
$$
\caption{SymTFT for twin gapped phases $\cP_i$ with symmetry $G^\omega$: the gapped physical boundary conditions are twin Lagrangian algebras, $\cL^1$ and $\cL^2$, so the same anyons $a$ can end. Nevertheless they will give rise to distinct order parameters $\cO_a^{(k)}$.  \label{fig:TwinSymTFT}}
\end{figure}

We identify a direct, stable second-order phase transition between the twin gapped phases. The transition is forced by the mixed-anomaly between the two subgroups $S_3^{(k)}$, as the symmetry generators of one broken symmetry carry charge under the other.
Crucially, 
gauging a symmetry $G$ does not alter the fact that the resulting gapped phases are twins, which means these are genuine {\bf phase transitions 
without any hidden spontaneous symmetry breaking} (SSB). The existence of these twin phases is tied to the existence of non-invertible symmetries with inequivalent full SSB phases.

\vspace*{1mm}
\noindent{\bf Intrinsic DQCPs are Twins.} More generally: any intrinsic DQCP transition in (1+1)d, i.e. without hidden symmetry breaking, comes from gapped phases that are twins: if the Lagrangian algebras for two gapped phases, with a DQCP transition are not twins, then placing one on the symmetry boundary of the SymTFT (corresponding to a specific gauging), results a full SSB for the same algebra and a partial SSB (or SPT) for the other.

\vspace*{1mm}
\noindent\textbf{Twin Phases via SymTFT.}
Let us briefly summarize the SymTFT for gapped phases \cite{Chatterjee:2022tyg,Moradi:2022lqp,Huang:2023pyk, Bhardwaj:2023fca, Bhardwaj:2023idu, Bhardwaj:2023bbf, Bhardwaj:2024qrf}. The SymTFT is a 2+1d TQFT on an interval with gapped symmetry boundary $\cL_\sym$, for a symmetry $\cS$, and gapped physical boundary $\cL_\phys$: after interval compactification, this is equivalent to a 1+1d $\cS$-symmetric gapped phase. 
Gapped boundary conditions of the SymTFT are given by Lagrangian algebras, which encode the anyons that end on the boundary. Lagrangian algebras decompose into anyons $a_i$ as
$\cL = \oplus_i n_i a_i$, $n_i\in \mathbb{N}$. 
Crucially however, $\cL$ carries an algebra structure. 
Two Lagrangian (or more genereally non-maximal condensable) algebras are {\bf twin algebras} if they agree in terms of the anyon decomposition, but differ in the algebra structures. 

This has the following important physical implication. 
Consider a SymTFT, with a fixed $\cL_\sym$, but two choices of physical boundary $\cL_\phys$ given by twin Lagrangian algebras 
$\cL_\phys^{k=1,2}$, Fig.~\ref{fig:TwinSymTFT}. This realizes twin gapped phases, irrespective of what the symmetry boundary is: 
the anyons that end on both boundaries, are local order parameters, anyons that only end on the physical boundary are string order parameters. However, due to the twin property of the physical boundaries, for each fixed symmetry, the general symmetry breaking type (by which we mean the number of ground states) of the two twin phases is the same. In particular, the twin phases never have relative SSBing. 

Twin phases are distinguished by distinct order parameters $\cO_a^{(1)}$ and $\cO_a^{(2)}$, which have the property that they come from distinct components of a
(higher-dimensional) anyon $a$, i.e. generalized charge $\Q_a$ \cite{Bhardwaj:2023ayw} in the SymTFT \footnote{The precise statement for general fusion categories is that the vector space of local operators is $\Hom_{\cZ(\cS)}(\cL_{\phys}, \cL_{\sym})=\oplus_{a\in \cL_{\phys}}n_a\Hom_{\cZ(\cS)}(a, \cL_{\sym})$. A local operator $\cO_a\in \Hom_{\cZ(\cS)}(a, \cL_{\sym})$.
}
\be
\Q_a \supsetneq  \cO_a^{(k)}  \,, \ \text{ for } k=1,2\,,\quad \cO_a^{(1)}\not= \cO_a^{(2)} \,.
\ee
A simple instance occurs for a non-abelian group symmetry $G$, with a higher-dimensional irrep $\rho$, and different components $r_k\subset \rho$ are the OPs for the twin phases.

\vspace*{1mm}
\noindent{\bf Twin Gapped BCs.} 
We consider the SymTFT of the simplest possible type: 
$D^\omega(G)$  finite $G$-gauge theory in 2+1d  with Dijkgraaf-Witten twist $\om\in H^3(G,U(1))$. The anyons are denoted by $([g], \rho)$, with conjugacy classes $[g]$ and irreps $\rho$ of the centralizer of $g$.
Gapped boundary conditions (i.e. Lagrangian algebras) of $D^\omega(G)$ are specified by a subgroup $H\subseteq G$ and $\gamma:G\times G\to U(1)$ such that $\om|_H=d\gamma$, {which are obtained from the Dirichlet boundary condition realizing $G^\omega$ symmetry by stacking with $\gamma$ and gauging $H$}. If $\om|_H\equiv 1$, then $\gamma\in H^2(H,U(1))$ is a 1+1d SPT. Examples of twin algebras have appeared in \cite{Pollmann:2012, Davydov:2013xov, Kobayashi:2025ykb, Kobayashi:2025pxs}, which are of the type $(H=G, \gamma_i)$, with different SPTs $\gamma_i$. In \cite{Gai:2026hjk} we provide explicit criteria and new constructions for twin algebras. Here we do not need $\gamma$, and label Lagrangians by the subgroups $H<G$
\be \label{eq:A_Lag}
    \cL_H:\qquad \text{Lagrangian obtained by gauging $H$} \,.
\ee
We consider twins based on subgroups $H, H'<G$, which are not conjugate in $G$ \footnote{Algebras labeled by non-conjugate subgroups are inequivalent. See \cite{Natale2017,Gai:2026hjk} for the complete equivalence relations.}
\be
\cL_1= \cL_H \,,\quad   \cL_2 = \cL_{H'} \,,\quad    H\not\sim_G H' \,.
\ee
 To be very concrete we will focus here on the simplest example, 
 which admits a simple direct phase transition between twin algebras, i.e. a no-hidden-SSB transition.
 
\vspace*{1mm}
\noindent\textbf{Preliminaries on $GL(2,3)$.} We will consider the order-48 group \footnote{In GAP \cite{GAP4} $G:=\text{SmallGroup}(48,29)$.} given by the general linear group on the 2d vector space over the field $\mathbb{F}_3$:
$G= GL(2,3) \cong Q_8 \rtimes S_3 $.
This group can also be faithfully represented as a subgroup of the {\bf single-qubit Clifford group $C_1$}, which also satisfies $C_1/\Z_8 \!=\! S_4\!=\!GL(2, 3)/\!\pm 1$. 
Denoting the generators of $GL(2,3)$ by: 
\be
GL(2,3)=  \langle xz, iz, c, h, a \rangle \,, 
\ee
 we can map into the Clifford group \footnote{For reference the standard Clifford gates are 
$
    H\!=\!\frac{1}{\sqrt{2}}
    \begin{pmatrix}
        1 & 1 \\ 1 & -1
    \end{pmatrix},\, 
    X\!=\!\begin{pmatrix}
        0 & 1 \\ 1 & 0
    \end{pmatrix},\, 
    P\!=\!\diag(1,i),\, Z\!=\!P^2
$} as follows
\be\ba
  xz&\mapsto XZ\,, & iz&\mapsto iZ, \quad c=(iz)^2=(xz)^2\mapsto-\bbI\cr 
    h&\mapsto H\,, &
    a&\mapsto e^{3\pi i/4} HPX \,.
\ea\ee
Define the following subgroups of $GL(2,3)=Q_8\rtimes S_3$
\be\label{SubGroupDef}
    S_3^{(1)}=\langle a,h \rangle\,,\quad 
    S_3^{(2)} =\langle a,hc \rangle\,,\quad 
    Q_8=\langle iz,\, xz\rangle\,.
\ee
Importantly, the non-normal $S_3^{(1)}$ and $S_3^{(2)}$ are not conjugate. Note that there is an outer automorphism that exchanges them. However, thinking of $GL(2,3)$ as a subgroup of $C_1$, the outer automorphism is conjugation by a non-Clifford gate and there is no {\it finite} group containing $G$, such that $S_3^{(1)}$ and $S_3^{(2)}$ are conjugate.

\vspace{2mm}
\noindent\textbf{Twin Gapped Phases for $GL(2,3)$.} The quantum double of $GL(2,3)$ admits twin Lagrangian algebras $\cL_{S_3^{(1)}}$ and $\cL_{S_3^{(2)}}$, with non-conjugate subgroups $S_3^{(k)}$ (\ref{SubGroupDef}), but whose anyon decomposition (``objects") is the same:
\be\ba\label{S3S3Twins}
\cL_{S_3^{(k)}} \cong_\obj 1 \oplus \rho_{3} \oplus \rho_{4} \oplus [h]_{++} \oplus [h]_{+-} \oplus [a]_{++} \oplus [a]_{+-} \,,
\ea\ee
where $\rho_n$ are $n$-dimensional irreps of $G$ (see App.~\ref{app:notations} for details of the anyons and App.~\ref{app:Hasse} for the Hasse diagram of algebras).  
Twin phases are obtained by placing the twin algebras (\ref{S3S3Twins}) on the physical boundary of the SymTFT. 
For $G$ symmetry, the twin phases, that we will denote by $\cP_{S_3^{(k)}}$ are partial SSB-phases, preserving $S_3^{(k)}$ for $k=1,2$ respectively. Their ground states correspond to cosets, labeled by the 8 elements of $Q_8$
\be \label{eq:cosets}
    {\GSr\;\; \longleftrightarrow\;\; S_3^{(k)} \, q,\qquad q\in Q_8} \,.
\ee
The phases are SSB phases with spontaneously broken $Q_8=\langle iz,xz \rangle$, however the preserved symmetries in each vacuum are distinct, and given by subgroups conjugate to $S_3^{(1)}$ and $S_3^{(2)}$, respectively. Crucially recall that the subgroups $S_3^{(k)}$ are not conjugate to each other, thus resulting indeed in distinct phases. 

Gauging non-abelian subgroups of $G$, we obtain twin phases with non-invertible symmetries: gauging $Q_8$ to $\cC(G, Q_8)$ results in {twin SPT} phases (which can also be obtained by gauging $\Z_8=\langle hiz\rangle$). {The transition between twin SPT phases is described by a single-universe CFT, also known as a gapless SPT (gSPT) phase.} Conversely, gauging $S_3^{(1)}$ to symmetry $\cC(G, S_3^{(1)})$ produces {twin full-SSB} phases. 
The three Lagrangian algebras for these symmetries are detailed in (\ref{app:Lags}) and the SymTFTs are shown in Fig.~\ref{fig:threephases}. Due to the twin algebra origin of the phases, however, there is no gauging that yields a relative SSB between the two gapped phases.

\vspace*{1mm}
\noindent
{\bf Twin Phase Transitions.}
Phase transitions in the SymTFT can be constructed by considering interfaces to reduced topological orders, determined by non-maximal, condensable algebras $\cA$ \cite{Frohlich:2003hm, davydov2013witt, Kong:2013aya}, which are contained in two Lagrangian algebras \cite{Chatterjee:2022tyg,Bhardwaj:2023bbf, Wen:2023otf, Bhardwaj:2024qrf}.

To obtain a direct, {\bf stable} transition between our twin gapped phases, we will generalize the setup to $G^\omega$, with $G= GL(2,3)$ and $\om\in H^3(G,U(1))$, with $\om|_{S_3^{(k)}}\equiv 1$, $k=1,2$, but a mixed anomaly between $h$ and $hc$ \footnote{The full 3-cocycle data, computed with GAP \cite{GAP4,HAP,mignard2017moritaequivalencepointedfusion,gruen2021computing} is provided in \texttt{GL23-om-data.g}.}:
\be\ba \label{eq:om}
    \om(h^{i_1}(hc)^{i_2},h^{j_1}(hc)^{j_2},h^{k_1}(hc)^{k_2})&=(-1)^{i_1j_1k_2}\,.
\ea\ee
The twin Lagrangian algebras continue to exist for $G^\omega$. For the physical boundary, those $\cL_H$ with $H$ containing the anomalous $\Z_2^{\langle c\rangle}$ are now absent. There are only partial and full twin-SSBs, no twin-SPTs. 
  
The non-maximal condensable algebras $\cA(H, N)$ relevant for the phase transition are labeled by a subgroups $H<G$ and $N\triangleleft H$ normal: $H$ is the preserved symmetry and $N$ acts trivially in the IR, in particular the charge anyons that restrict trivially to $H$ and the group elements of $N$ can end on the interface. 
Our twin phase transition is obtained from the subgroups
\be\ba
H = \langle S_3^{(1)}, S_3^{(2)}\rangle = \langle a,h,hc\rangle \cong D_{12}\,,\quad 
N= \Z_3^{\langle a \rangle}  \,,
\ea\ee
and has anyon decomposition 
\be\label{AHN}
\cA_{(H, N)} \cong_\obj 1 \oplus \rho_{3} \oplus [a]_{++}\,.
\ee
Importantly, $H$ is now anomalous: this rules out gapped phases with $H=\langle S_3^{(1)}, S_3^{(2)}\rangle$ symmetry that would allow the two consecutive Landau transitions $S_3^{(1)} < H$ and $H>S_3^{(2)}$. 
Thanks to the anomaly, the gapped phase with $H$ symmetry is now excluded and we will show how to construct a stable transition directly between the twin phases preserving $S_3^{(1)}$ and $S_3^{(2)}$.

The reduced TO for $\cA_{(H, N)}$ is $D^{\omega}(H/N)=D^{\om_\II}(\Z_2^2)$, whose anyons we denote by $m_1,e_1,m_2,e_2$. Note that because of the mixed anomaly, the (bosonic) magnetic anyon for one symmetry carries a fractionalized charge (mathematically, a projective representation) for the other symmetry, as can be seen from the fusions $m_1^2=e_2,\;m_2^2=e_1$. Furthermore, the anyons carrying a flux of $m_1m_2$ for the anomalous diagonal $\Z_2$ have topological spin $\pm i$. The interface specified by $\cA_{(H, N)}$ corresponds to a gapped boundary of $D^\om(G)\boxtimes\ol{D^{\om_\II}(\Z_2^2)}$ labeled by the subgroup 
\be \label{eq:Hdiag}
    H^\diag=\langle (h,m_1), (hc,m_2), (a,\id) \rangle \subset G\times\Z_2^2\,.
\ee
Note that, since $h$ and $hc$ are in the same $G$-conjugacy class $[h]$, both anyons $m_1$ and $m_2$ will be mapped to the same anyon $[h]_{++}$. Similarly, the electric anyons $e_1$ and $e_2$ are both mapped to $\rho_4$:
\be \label{eq:interface_map_short}
\begin{split}
\begin{tikzpicture}
\begin{scope}[shift={(0,0)}]
\draw [SymTFTColores, fill= SymTFTColores, opacity = 0.2] 
(0,-0.2) -- (0,2) -- (2,2) -- (2,-0.2) -- (0,-0.2) ; 
\draw [very thick] (0,-0.2) -- (0,2) ;
\draw [very thick] (2,-0.2) -- (2,2) ;
\draw [ thick] (0,0.8) -- (4,0.8) ;
\draw [ thick] (0,0.2) -- (4,0.2) ;
\fill[] (0,0.8) circle (0.05cm);
\fill[] (4,0.8) circle (0.05cm);
\fill[] (0,0.2) circle (0.05cm);
\fill[] (4,0.2) circle (0.05cm);
\node[above] at (1,0.8) {$\rho_4$};
\node[above] at (1,0.2) {$[h]_{++}$};
\node[above] at  (1,1.3)  {$D^\om(G)$} ;
\node[above] at (0,2) {$\cL_\sym^{G,\om}$}; 
\node[above] at (2,2) {$\cA(H, N)$}; 
\end{scope}
\begin{scope}[shift={(2,0)}]
\draw [SymTFTColores, fill= SymTFTColores, opacity = 0.1] 
(0,-0.2) -- (0,2) -- (2,2) -- (2,-0.2) -- (0,-0.2) ; 
\draw [very thick] (2,-0.2) -- (2,2) ;
\node[above] at  (1,1.3)  {$D^{\om_{\II}}(\Z_2^2)$} ;
\node[above] at (1,0.8) {$e_1,\;e_2$};
\node[above] at (1,0.2) {$m_1,\;m_2$};
\node[above] at  (2,2)  {$\cL_{\phys}$} ;
\end{scope}
\end{tikzpicture}
\end{split}
\ee 
The complete map is provided in \eqref{eq:FoldedLag}.
The Lagrangian algebras of $D^{\om_{\II}}(\Z_2^2)$ that get mapped to the twin algebras in $D^\om(G)$ are:
\be\ba \label{eq:RedTO_Lags}
        \cL_{\langle m_1\rangle} & \cong  1 \oplus m_1 \oplus e_2 \oplus m_1e_2 &&\mapsto\;\; \cL_{S_3^{(1)}}\\
        \cL_{\langle m_2\rangle} & \cong 1 \oplus m_2 \oplus e_1 \oplus m_2e_1 &&\mapsto\;\;\cL_{S_3^{(2)}}\,,
\ea\ee
which we label by the preserved subgroup of $\Z_2^2$.
By imputing a $(\Z_2^2)^{\om_\II}$ symmetric phase transition $\CFT^{\Z_2^2,\om^\II}$ (e.g. the $c=1$ compact boson CFT with mixed-anomaly between shift and winding symmetries \cite{Thorngren:2021yso}) and mapping through the SymTFT interface \cite{Bhardwaj:2023bbf,Bhardwaj:2024qrf}, we obtain a second order phase transition between the twin phases $\cP_{S_3^{(1)}}$ and $\cP_{S_3^{(2)}}$ for the group $G=GL(2,3)$ with anomaly $\omega$:
\be\ba \label{eq:GomCFT}
\begin{tikzpicture}
 \begin{scope}[shift={(0,0)}, local bounding box=yourbox]
\node (p1) at (1,0) {$\CFT^{G,\om}$};
\node (p2) at (-1.5,-0.5) {$\cP_{S_3^{(1)}}$};
\node (p3) at (3.5,-0.5) {$\cP_{S_3^{(2)}}$};
 \draw [-stealth] (p1) edge (p2);
 \draw [-stealth] (p1) edge (p3);
 \node at (-0.3,0) {$\cO_{\rho_4}^{(1,1)}$};
 \node at (2.3,0) {$\cO_{\rho_4}^{(2,2)}$};
 \end{scope}
 \node [draw= SymTFTColores, fill= SymTFTColores, opacity = 0.2, inner sep=1pt, fit=(yourbox)] {};
 \node at (2.4,-1.2) {$\Uparrow$ Map via $\cA (H, N)$};
  \begin{scope}[shift={(0,-1.8)},local bounding box=mybox]
\node (p1) at (1,0) {$\CFT^{\Z_2^2,\om^\II}$};
\node (p2) at (-1.5,-0.5) {$\cP_{\langle m_1\rangle}$};
\node (p3) at (3.5,-0.5) {$\cP_{\langle m_2\rangle}$};
 \draw [-stealth] (p1) edge (p2);
 \draw [-stealth] (p1) edge (p3);
 \node at (-0.3,0) {$\cO_{e_2}$};
 \node at (2.3,0) {$\cO_{e_1}$};
 \end{scope}
\node [draw=SymTFTColores, fill= SymTFTColores, opacity = 0.1, inner sep=1pt, fit=(mybox)] {};
\end{tikzpicture}
\ea
\ee
Notice that the $\Z_2^2$ OPs $\cO_{e_2}$ and $\cO_{e_1}$ are mapped to different components of the same $G$-irrep $\rho_4$ \footnote{There are also other OPs coming from $\rho_3$, but these are identical in both phases.}. The particular component gaining a vev can be computed from the explicit matrices \eqref{eq:rho4}, which show that $\rho_4|_{S_3^{(k)}}$ decomposes into $S_3$ irreps as:
\be
    \rho_4|_{S_3^{(1)}}= 1\oplus \rho_{1-} \oplus \rho_2\,,\quad 
    \rho_4|_{S_3^{(2)}}= \rho_{1-} \oplus 1 \oplus \rho_2\,.
\ee
The component that has a vev in $\cP_{S_3^{(k)}}$ is the trivial $S_3$ irrep $1$, which appears in entry $(k,k)$.  The symmetry $H=\langle S_3^{(1)}, S_3^{(2)}\rangle =D_{12}$ is anomalous and the gapless phase $\cP_\CFT^{G,\om}$ is an intrinsically gapless SSB (igSSB) phase \cite{Bhardwaj:2024qrf,Warman:2024lir}: it has four vacua, each with $\CFT^{\Z_2^2,\om^\II}$ (e.g the compact boson) preserving ${D_{12}}$ and related by the spontaneously broken $Q_8$ generators $g={iz},{xz}$.
When we deform the gapless phase by $\cO_{\rho_4}^{(k,k)}$ and gap the theory, each vacuum splits into two resulting in the twin gapped phases with 8 vacua \eqref{eq:cosets}.

The $(\Z_2^2,\om_\II)$ DQCP transition can be mapped to a Landau $\Z_4$ transition \cite{Zhang:2022wwn}. From the SymTFT perspective, this is obtained by changing the symmetry boundary to $\cL_{\langle m_1\rangle}$. However, this is not the case for our transition: after mapping it through the interface $\cA(H, N)$, we obtain a non-Landau continuous phase transition between twin full SSBs for the non-invertible symmetry $\cS=\cC(G,S_3^{(1)})$:
\be\ba \label{eq:cSCFT}
\begin{tikzpicture}
 \begin{scope}[shift={(0,0)}, local bounding box=yourbox]
\node (p1) at (1,0) {$\CFT^{\cS}$};
\node (p2) at (-1.5,-0.5) {Full $\cS$-SSB 1};
\node (p3) at (3.5,-0.5) {Full $\cS$-SSB 2};
 \draw [-stealth] (p1) edge (p2);
 \draw [-stealth] (p1) edge (p3);
 \end{scope}
 \node [draw=SymTFTColores, fill= SymTFTColores, opacity = 0.2, inner sep=2pt, fit=(yourbox)] {};
  \node at (2.3,-1.1) {$\Uparrow$ Map via $\cA (H, N)$};
  \begin{scope}[shift={(0,-1.7)},local bounding box=mybox]
\node (p1) at (1,0) {$\CFT^{\Z_4}$};
\node (p2) at (-1.5,-0.5) {$\Z_4$ SSB};
\node (p3) at (3.5,-0.5) {$\Z_4$ Trivial};
 \draw [-stealth] (p1) edge (p2);
 \draw [-stealth] (p1) edge (p3);
 \end{scope}
\node [draw=SymTFTColores, fill= SymTFTColores, opacity = 0.1, inner sep=2pt, fit=(mybox)] {};
\end{tikzpicture}
\ea
\ee
In conclusion, the $\CFT^{G,\om}$ and $\CFT^{\cS}$ that we construct in \eqref{eq:GomCFT} and \eqref{eq:cSCFT} are second-order phase transitions between twin gapped phases. Furthermore, as we already discussed in generality, and shown explicitly for this example, twin phases will never have relative SSBs, even after (partially) gauging some symmetry, so they are intrinsically beyond Landau transitions.

\begin{figure}[t]
	\centering
	\includegraphics[width=0.9\linewidth]{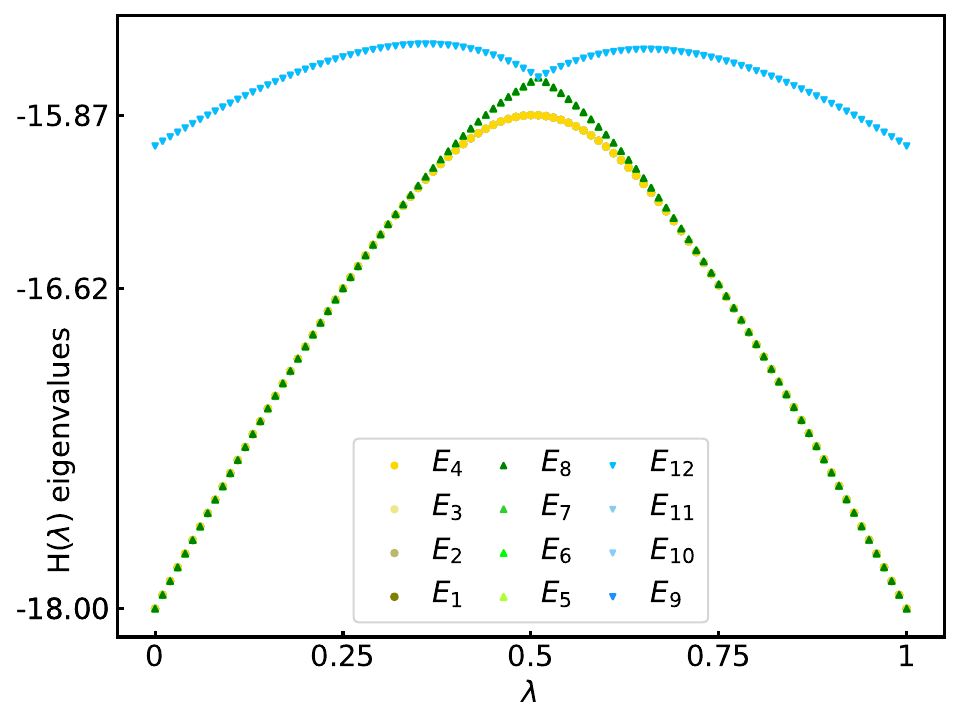}
    \hspace*{4mm}\includegraphics[width=0.87\linewidth]{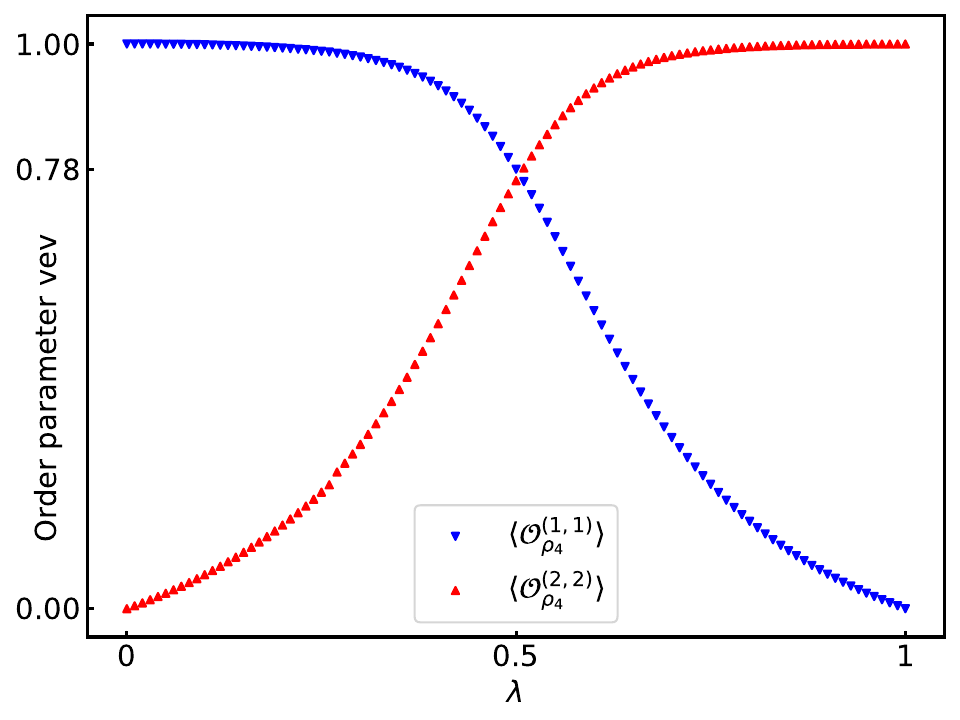} \vspace*{-3mm}
	\caption{Characterization of the intrinsic DQCP quantum phase transition for $L=3$ sites. 
    Top: the lowest 12 eigenvalues of $H(\lambda)$. The twin gapped phases at $\lambda=0,1$ have $8$ degenerate ground states, while the critical point at $\lambda=1/2$ has $4$. The transition is continuous (second-order).  
    Bottom: the vev of $\cO_{\rho_4}^{(k,k)}=\frac{1}{L}\!\!\sum\limits_{j=1,\cdots,L}\left(Z^{\rho_4}_{j}\cdot Z^{\rho_4\dagger}_{j+1}\right)_{k,k}$ for $k=1,2$.}
	\label{fig:plots}
    \vspace*{-3mm}
\end{figure}

\vspace*{1mm}
\noindent{\bf Lattice Model.}
To corroborate our findings, we consider the lattice model based on the anyon chain for $GL(2,3)^\omega$. Detailed derivations are given in App.~\ref{app:Lattice}. Consider a periodic chain on a spatial circle whose local Hilbert space on each edge has basis $\{\ket{f}\;:\;f\in G\}$. 
The symmetry action by $g\in G$ comprises of right multiplication and the anomalous phase $\omega(g_{j}g_{j+1}^{-1},g_{j+1},g)$ between all neighboring sites: \\[-5mm]
\be\ba 
    U_g\ket{g_1,\cdots,g_L}\!=\!\prod_j \omega(g_{j}g_{j+1}^{-1},g_{j+1},g)\ket{g_1g,\cdots,g_Lg}\,.\\[-3mm]
\ea\ee
The Hamiltonians contain projectors that SSB $G\to S_3^{(k)}$, written from diagonal operators
\be 
    Z^{\rho}_{n,m}\ket{g}=
    \rho(g)_{n,m}\ket{g}\,,
\ee
(where $\rho(g)$ is the unitary matrix representation of $g$ in the irrep $\rho$, and $n,m$ label its row and column indices) and disordering operators for $f\in S_3^{(k)}$ \\[-3mm]
\be\ba 
        L_{j+1}^{f,\om}\ket{g_{j},\,g_{j+1},\,g_{j+2}}=
        &\frac{\omega(g_{j}g_{j+1}^{-1}f^{-1},\;f,\;g_{j+1})}{\omega(f,\;g_{j+1}g_{j+2}^{-1},\;g_{j+2})}\times\\
        &\ket{g_{j},\,fg_{j+1},\,g_{j+2}}\,. \\[-2mm]
\ea\ee
The interpolation $H(\lambda)$ between the Hamiltonians $H_{S_3^{(k)}}$ for the twin gapped phases $k=1,2$ is: \footnote{We add the Hermitian Conjugate (h.c.) of the operators, so $H(\lambda)$ is Hermitian.}
\begin{align} \label{eq:Hlambda_main}
    H(\lambda)=\,&(1-\lambda)H_{S_3^{(1)}}+\lambda H_{S_3^{(2)}} \nn \\
    \simeq&-\sum_j\frac{1}{8}\left[\mathbb I_{j,j+1}\!
    +3\!\left(Z^{\rho_3}_{j}\cdot Z^{\rho_3\dagger}_{j+1}\right)_{1,1}\right]_j \nn\\
    &-\frac{1}{3}\sum_j\lb\bbI_j+L_j^{a,\om}+L_j^{a^2,\om}\rb -\frac{1}{2}\sum_j\bbI_j  \\
    &-\frac{1}{2}\sum_j (1-\lambda)\lbb L_j^{h,\om}+\left(Z^{\rho_4}_{j}\cdot Z^{\rho_4\dagger}_{j+1}\right)_{1,1}\rbb \nn\\
    &-\frac{1}{2}\sum_j \lambda\lbb L_j^{hc,\om}+\left(Z^{\rho_4}_{j}\cdot Z^{\rho_4\dagger}_{j+1}\right)_{2,2}\rbb+\text{h.c.} \nn
\end{align}
which shows that the operator gaining a vev is a different component of $\rho_4$ for the two twin phases $k=1,2$: 
\be\ba
    \GSkl \left(Z^{\rho_4}_{j}\cdot Z^{\rho_4\dagger}_{j+1}\right)_{k,k} \GSkr=1\,,
\ea\ee

The phase transition occurs at $\lambda=1/2$, see Fig.~\ref{fig:plots} for numerical confirmation with plots of the eigenvalues of $H(\lambda)$ and order parameter vevs.

\twocolumngrid
\vspace{2mm}
\noindent
{\bf Necessary Conditions.}
A natural question is what should replace (\ref{Landau}). We can formulate this for a general symmetry category $\cS$, with Drinfeld center (i.e. category of anyons in the SymTFT) $\cZ(\cS)$. What is the analog of the preserved symmetry of an $\cS$-symmetric gapped phase, with physical boundary given by the Lagrangian algebra $\cL$? Let $F_\cS: \cZ (\cS) \to \cS$ be the forgetful functor \cite{davydov2013witt}. Then the 
\be
\text{preserved symmetry is}\   F_\cS (\cL)  \,,
\ee
which forms a Frobenius algebra in $\cS$ \cite{Frohlich:2003hm, EGNO, Bhardwaj:2023ayw}. These are generally not sub-categories but algebras in $\cS$. 
The condition for relative SSB-ing in (\ref{Landau}) between gapped phases with physical boundary condition given by $\cL_1$ and $\cL_2$ becomes one of Frobenius sub-algebras:
\be\label{FSL}
F_\cS (\cL_1) < F_\cS (\cL_2) \,.  
\ee
The existence of hidden symmetry breaking means that there exists an $\cS'$, Morita equivalent (i.e.~related by gauging) to $\cS$, such that (\ref{FSL}) holds for $\cS'$. 
However, this is not the most general criterion, as it misses the twin phases. 
The natural conjecture from the analysis of twin algebras is the following criterion: let $\cL_1$ and $\cL_2$ be two inequivalent gapped physical boundary conditions. Then automatically, $F_\cS(\cL_k)$ are inequivalent as Frobenius algebras. A necessary condition for a {stable} second order transition is then that there does not exist a Lagrangian algebra $\cL_3$ such that 
\be
F_{\cS} (\cL_3) > F_\cS (\cL_k) \,,\qquad \text{for }k=1,2  \,.
\ee
A candidate {direct} second order transition is obtained by considering the largest condensable algebra $\cA_{12}$ in $\cZ(\cS)$ that is contained in both $\cL_{1}$ and $\cL_{2}$, using the club sandwich construction. 
To conclude let us illustrate our findings in this language: $\cS= GL(2,3)^{\omega}$. The two Lagrangians are the twins $\cL_{S^{(k)}_3}$. Without $\omega$ we have an $\cL_3=\cL_{D_{12}}$ \footnote{The $\cL_3$ is given by 
$
\cL_{D_{12}}=1 \oplus \rho_3 \oplus 2[h]_{++} \oplus [a]_{++} \oplus [c] \oplus [c]\rho_3 \oplus [ca]$.
}
that allows a composition of two Landau transitions $S_3^{(1)}< D_{12}$ and $D_{12}> S_3^{(2)}$. With anomaly this is not a Lagrangian algebra and the candidate condensable algebra for the transition is $\cA_{12}= \cA (H, N)$ in (\ref{AHN}). It is the largest condensable algebra that sits above the two Lagrangians in the Hasse diagram Fig.~\ref{fig:HasseG}, which indeed furnished the transition in (\ref{eq:GomCFT}). A more in depth characterization of transitions including DQCPs through the structure of condensable algebras will be discussed in \cite{SWW}.

\onecolumngrid
\noindent
\begin{acknowledgments}
\vspace{-3mm}
We thank Andrea Antinucci, Thomas Bartsch, Christian Copetti, Andr\'{e} Henriques, Kansei Inamura, Ryohei Kobayashi, John McGreevy, Kantaro Ohmori, Sal Pace, Alex Turzillo,  Rui Wen, Minyoung You, Yunqin Zheng for discussions. We are grateful to Rui Wen for comments on a draft {and to Kansei Inamura for helpful advice on the lattice model}. SSN thanks the anonymous referee(s?) who over the years have kept insisting in their reports on asking about the algebra structure of condensable algebras, for sparking this exciting exploration. AW thanks Xiao-Gang Wen for suggesting the computation algebra system GAP \cite{GAP4}. This work is supported by the UKRI Frontier Research Grant, underwriting the ERC Advanced Grant ``Generalized Symmetries in Quantum Field Theory and Quantum Gravity''. The data that support the findings of this article are openly available \cite{data}.
\end{acknowledgments}


\twocolumngrid
\let\oldaddcontentsline\addcontentsline
\renewcommand{\addcontentsline}[3]{}
\bibliographystyle{ytphys}
 \let\bbb\bibitem\def\bibitem{\itemsep1.6pt\bbb}
\bibliography{GenSym.bib}
\let\addcontentsline\oldaddcontentsline


\clearpage
\newpage
\appendix
\section{Details on the SymTFT for $GL(2,3)$}
\label{app:notations}

In this appendix we provide details of the SymTFT, anyons, algebras and other structures related to the example of $G= GL(2,3)$.

\subsection{Anyons and Algebras}
\noindent\textbf{Notation for Anyons.} We denote anyons in $D^\omega(G)$ by 
$([g],\rho_g)$, where $[g]=\{fgf^{-1}: f\in G\}$ is a conjugacy class of $G$ and $\rho_g$ is a (projective) representation of the centralizer 
$C_G(g)=\{f\in G\;|\;gf=fg\}\,.$

Let us first focus on anyons with trivial flux $g=\id$, for which $C_G(\id)=G$ and $\rho$ is a representation of $G$. The characters (i.e. the traces of the representation matrices) of the $G$-irreps $\rho_n$ are shown in Tab.~\ref{tab:chGL23}. 

We fix the following matrices for the unfaithful irrep $\rho_3:G\to U(3)$, where $\omega=e^{2\pi i/3}$ and $\rho_3(c)=\bbI$
\be\ba 
\rho_3(h)\;=\;
&\begin{pmatrix}
1 & 0 & 0 \\
0 & 0 & 1 \\
0 & 1 & 0 \\
\end{pmatrix},\; &
\rho_3(xz)=\frac{1}{3}
&\begin{pmatrix}
-1 & 2\omega & 2\omega \\
2\omega^2 & -1 & 2 \\
2\omega^2 & 2 & -1 \\
\end{pmatrix}, \\
\rho_3(a)\;=\;
&\begin{pmatrix}
1 & 0 & 0 \\
0 & \omega & 0 \\
0 & 0 & \omega^2 \\
\end{pmatrix}, &
\rho_3(iz)=\frac{1}{3}
&\begin{pmatrix}
-1 & 2 & 2\omega^2 \\
2 & -1 & 2\omega^2 \\
2\omega & 2\omega & -1 \\
\end{pmatrix},
\ea\ee
and for the faithful irrep $\rho_4:G\to U(4)$
\be\ba \label{eq:rho4}
\rho_4(h) &=
\begin{pmatrix}
1 & 0 & 0 & 0 \\
0 & -1 & 0 & 0 \\
0 & 0 & 0 & 1 \\
0 & 0 & 1 & 0 \\
\end{pmatrix},  
\quad \rho_3(a)=
\begin{pmatrix}
1 & 0 & 0 & 0 \\
0 & 1 & 0 & 0 \\
0 & 0 & \omega & 0 \\
0 & 0 & 0 & \omega^2 \\
\end{pmatrix},
\\
\rho_4(xz)&=\frac{1}{\sqrt{3}}
\begin{pmatrix}
 0 & -i & i & -i \\
 -i & 0 & -i & -i \\
 i & -i & -i & 0 \\
 -i & -i & 0 & i \\
\end{pmatrix}, \qquad \rho_4(c)=-\bbI\cr 
\rho_4(iz)&=\frac{1}{\sqrt{3}}
\begin{pmatrix}
0 & i & e^{i\pi 5/6} & e^{i\pi 7/6} \\
i & 0 & -e^{i\pi 5/6} & e^{i\pi 7/6} \\
-e^{i\pi 7/6} & e^{i\pi 7/6} & i & 0 \\
-e^{i\pi 5/6} & -e^{i\pi 5/6} & 0 & -i \\
\end{pmatrix}.
\ea\ee

We now turn to anyons $([g],\rho_g)$ with non-trivial flux $[g]$. If $\omega$ is non-trivial,  the representation of the centralizer $C_G(g)$ is projective, i.e.
\be
    \rho_g(g_1)\rho_g(g_2)=i_g\om(g_1,g_2)\,\rho_g(g_1g_2)
\ee
where $i_g\om$ is determined from the anomaly $\omega$ \cite{Dijkgraaf:1989pz,Coste:2000tq,gruen2021computing}:
\be
    i_g\om(g_1,g_2)=\frac{\omega(g, g_1, g_2)\omega(g_1, g_2, g)}{\omega(g_1, g, g_2)}\,.
\ee
Let us discuss the anyons with non-trivial flux appearing in the twin algebras $\cL_{S_3^{(k)}}$. 
First note that $C_G(a)=\Z_3^{\langle a\rangle}\times\Z_2^{\langle c\rangle}$,  $C_G(h)=\Z_2^{\langle h\rangle}\times\Z_2^{\langle c\rangle}$ and that with our choice of 3-cocycle, $i_a\om\equiv1$,  $i_h\om\equiv1$. The irreps $++$ and $+-$ map $a,h$ to $+1$ and map $c$ to $+1$ and $-1$, respectively. Note that for the anyons with flux in $[h]$, we could also have chosen $hc\in[h]$ to be its representative group element. In this case, 
$i_{hc}\om(hc,hc)=+1$ but $i_{hc}\om(h,h)=-1$
so the bosons with $hc\mapsto1$ will carry projective representations in which $h\mapsto\pm i$: therefore the $hc$ fluxes carry fractionalized charge under $h$ (and conversely, if we change representative 3-cocycle) \footnote{The characters of $C_G(hc)$ can also be computed from those of $C_G(h)$ by means of \cite[equation (2.10)]{gruen2021computing}.}.   

\vspace*{1mm}
\noindent\textbf{Symmetry Lagrangian Algebras.} We consider three symmetry boundaries: 
\be\label{app:Lags}
\ba
G:\quad \cL_\sym^G\cong_\obj &\oplus_{\rho\in \text{Irr}(G)} \dim(\rho)\,\rho \cr 
\cC(G, Q_8) :\quad  
   \cL_{Q_8} \cong_\obj 
   &1 \oplus  \rho_{1-} \oplus 2 \rho_{2} \oplus [c] \oplus [c]\rho_{1-}  \cr &  \oplus 2[c]\rho_{2} \oplus 3[iz]_1 \oplus 3[iz]_{-1}\cr 
\cC(G,S_3^{(1)}) :\quad    
\cL_{S_3^{(1)}} \cong_\obj &1 \oplus \rho_{3} \oplus \rho_{4} \oplus [h]_{++} \oplus [h]_{+-} \\[-1mm]
&\oplus [a]_{++} \oplus [a]_{+-} \,
\ea
\ee
The notation $\cC(G,K)$ is the symmetry category obtained after gauging $K$ in $G$, which  for non-abelian groups contains non-invertible objects \cite{Ostrikmodule}. For flux $[iz]$, we have $C_G(iz)={\langle hxza\rangle}\cong\Z_8$: without anomaly its generator carries $+1$ and $-1$ irrep for $[iz]_1$ and $[iz]_{-1}$ respectively. Since $(hxza)^2=iz$ both these anyons are bosons. Note that in the case with mixed anomaly $\om$ considered in the main text, $Q_8$ can no longer be gauged, these anyons will no longer be bosons and there will be no corresponding $\cC(G,Q_8)$ Lagrangian algebra.

\begin{table}[t]
$$
{\scriptsize\begin{array}{|l|cccccccc|}
\hline
        &  [1] & [c] & [a] & [ac] & [h] & [iz] & [hiz] & [h(iz)^3] \\
\hline
  \;\;1  & 1 & 1 & 1 & 1 & 1 & 1 & 1 & 1 \\
  \rho_{1_-}  & 1 & 1 & 1 & 1 & -1 & 1 & -1 & -1 \\
  \rho_{2}  & 2 & 2 & -1 & -1 & 0 & 2 & 0 & 0 \\
  \rho_{2_+}  & 2 & -2 & -1 & 1 & 0 & 0 & \sqrt{2}\,i & -\sqrt{2}\,i \\
  \rho_{2_-}  & 2 & -2 & -1 & 1 & 0 & 0 & -\sqrt{2}\,i & \sqrt{2}\,i \\
  \rho_{3}  & 3 & 3 & 0 & 0 & 1 & -1 & -1 & -1 \\
  \rho_{3_-}  & 3 & 3 & 0 & 0 & -1 & -1 & 1 & 1 \\
   \rho_{4} & 4 & -4 & 1 & -1 & 0 & 0 & 0 & 0 \\
\hline
\end{array}
}
$$
\caption{Character table for the group $GL(2,3)$. \label{tab:chGL23}
}
\end{table}

\vspace*{1mm}
\noindent{\bf Map of Anyons.} This is computed from the decomposition of a Lagrangian algebra in $D^\om(G)\boxtimes\ol{D^{\om_\II}(\Z_2^2)}$ labeled by the subgroup 
\be 
    H^\diag=\langle (h,m_1), (hc,m_2), (a,\id) \rangle \subset G\times\Z_2^2\,.
\ee
into anyons, (see \cite{Gai:2026hjk,delaFuente:2023whm} for the formula)
\be\ba\label{eq:FoldedLag}
1  &\mapsto  1 \oplus \rho_{3} \oplus [a]_{++}   \\
e_1\,,\; e_2  &\mapsto  \rho_{4} \oplus [a]_{+-}   \\
e_1 e_2  &\mapsto  \rho_{1-} \oplus \rho_{3-} \oplus [a]_{++}   \\
m_1\,,\; m_2  &\mapsto  [h]_{++}   \\
m_1 e_2\,,\; m_2 e_1 &\mapsto  [h]_{+-}   \\
m_1 e_1\,,\; m_2 e_2 &\mapsto  [h]_{-+}   \\
m_1 e_1 e_2\,,\; m_2 e_1 e_2 &\mapsto  [h]_{--}   \\
m_1 m_2   &\mapsto  [c]_{-i} \oplus [c]_{-i}\rho_{3} \oplus [ac]_{+-i}  \\
m_1 m_2 e_1 e_2  &\mapsto  [c]_{-i}\rho_{1-} \oplus [c]_{-i}\rho_{3-} \oplus [ac]_{+-i}   \\
m_1 m_2 e_1\,,\; m_1 m_2 e_2  &\mapsto  [c]_{-i}\rho_4 \oplus [ac]_{++i}   \\
\ea\ee
Note that, in our conventions, $m_1m_2$ and $m_1m_2e_1e_2$, and all the anyons they map to, have topological spin $-i$. Indeed, due to the anomaly $\omega$, the anyon in $D^\om(G)$ with flux $[c]$ carries a projective representation of $G$ in which $c\mapsto-i$, which is why we denote it as $[c]_{-i}$. $m_1 m_2 e_1$ and $m_1 m_2 e_2$ instead have spin $i$ as do $[c]_{-i}\rho_4$ (since $\rho_4:c\mapsto -1$ so changes the spin by a sign) as does $[ac]_{++i}$.

\subsection{Symmetry Categories from Gauging $Q_8$ or $S_3$} \label{app:symm_cats}

\begin{figure}
\centering
$$
\begin{tikzpicture}
\begin{scope}[shift={(0,0)}]
\draw [SymTFTColores,  fill=SymTFTColores, opacity = 0.2] 
(0,0) -- (0,3) -- (3,3) -- (3,0) -- (0,0) ; 
\draw [white] (0,0) -- (0,3) -- (3,3) -- (3,0) -- (0,0)  ; 
\draw [very thick] (0,0) -- (0,3) ;
\draw [very thick] (3,0) -- (3,3) ;
\draw [thick](3,2.1) -- (0,1.8);
\draw [thick](3,2.1) -- (0,2);
\draw [thick](3,2.1) -- (0,2.2);
\draw [thick](3,2.1) -- (0,2.4);
\draw [thick](3,1) -- (0,1.3);
\draw [thick] (3,1) -- (0,1);
\draw [thick] (3,1) -- (0,0.7);
\draw [thick] (0, 0.3) -- (3,0.3);
\fill[] (3,2.1) circle (0.05cm);
\fill[] (3,1) circle (0.05cm);
\fill[] (3,0.3) circle (0.05cm);
\fill[] (0,1.8) circle (0.05cm);
\fill[] (0,2) circle (0.05cm);
\fill[] (0,2.2) circle (0.05cm);
\fill[] (0,2.4) circle (0.05cm);
\fill[] (0,1.3) circle (0.05cm);
\fill[] (0,1) circle (0.05cm);
\fill[] (0,0.7) circle (0.05cm);
\fill[] (0,0.3) circle (0.05cm);
\node[above] at (1.5, 2.3) {$\rho_4$};
\node[above] at (1.5, 1.2) {$\rho_3$};
\node[above] at (1.5, 0.3) {$1$};
\node[above] at  (1.5,3)  {$D(G)$} ;
\node[above] at (0,3) {$G$}; 
\node[above] at (3,3) {$\cL_{S_3^{(k)}}$}; 
\end{scope}
\begin{scope}[shift={(0,-1.7)}] 
					\draw [SymTFTColores, fill=SymTFTColores, opacity = 0.2]
					(0,0) -- (0,1) -- (3,1) -- (3,0) --(0,0);
					\draw[very thick] (0,0) -- (0,1) ;
                    \draw[very thick] (3,0) -- (3,1) ;
					\draw[thick] (0,0.5) -- (3,0.5);
                    \fill[] (0,0.5) circle (0.05cm);
                     \fill[] (3,0.5) circle (0.05cm);
					\node[above]  at  (3,1) {$\cL_{S_3^{(k)}}$};
					\node[above]  at (0,1) {$\cC(G, Q_8)$};
					\node[above]  at (1.5, 0.5) {$1$};
	                   \end{scope}
\begin{scope}[shift={(4.5,-1)}] 
\draw [SymTFTColores, fill=SymTFTColores, opacity = 0.2]
(0,0) -- (0,4) -- (3,4) -- (3,0) -- (0,0) ; 
\draw [white] (0,0) -- (0,4) -- (3,4) -- (3,0) -- (0,0)  ; 
\draw [very thick] (0,0) -- (0,4) ;
\draw [very thick] (3,0) -- (3,4) ;
\draw[thick] (0,0.5) -- (3,0.5);
\draw[thick] (0,1) -- (3,1);
\draw[thick] (0,1.5) -- (3,1.5);
\draw[thick] (0,2) -- (3,2);
\draw[thick] (0,2.5) -- (3,2.5);
\draw[thick] (0,3) -- (3,3);
\draw[thick] (0,3.5) -- (3,3.5);
 \fill[] (0,1) circle (0.05cm);
  \fill[] (0,1.5) circle (0.05cm);
   \fill[] (0,0.5) circle (0.05cm);
    \fill[] (0,2) circle (0.05cm);
     \fill[] (0,2.5) circle (0.05cm);
      \fill[] (0,3) circle (0.05cm);
       \fill[] (0,3.5) circle (0.05cm);
        \fill[] (0,1) circle (0.05cm);
  \fill[] (3,1.5) circle (0.05cm);
   \fill[] (3,0.5) circle (0.05cm);
    \fill[] (3,2) circle (0.05cm);
     \fill[] (3,2.5) circle (0.05cm);
      \fill[] (3,3) circle (0.05cm);
       \fill[] (3,3.5) circle (0.05cm);
					\node[above]  at (1.5, 0.5) {$1$};
					\node[above]  at (1.5, 1)  {$\rho_3$};
                    \node[above]   at (1.5, 1.5)  {$\rho_4$};
				    \node[above]  at (1.5, 2)  {$[h]_{++}$};
					\node[above]   at(1.5, 2.5)   {$[h]_{+-}$};
                   \node[above]   at (1.5, 3)  {$[a]_{++}$}; 
					\node[above]   at (1.5, 3.5)  {$[a]_{+-}$};
					\node[above]  at  (3,4) { $\cL_{S_3^{(k)}}$};
					\node[above]  at (0,4) {$\cC(G, S_3^{(1)})$};
                    \end{scope}
			\end{tikzpicture}    
$$
\caption{Three twin gapped phases: 
the physical boundaries are the twin Lagrangian algebras $\cL_{S_3^{(1)}}$ and $\cL_{S_3^{(2)}}$. Changing the symmetry boundary results in different overall symmetry breaking patterns, but no relative SSB-ing between the twin phases. Clockwise: 
$\cS$ is the $G$-group symmetry (twin partial SSB phase), gauging $S_3^{(1)}$ gives a non-invertible $\cC(G, S_3^{(1)})$ symmetry (twin full SSB phases) and and the non-invertible $\cC(G, Q_8)$ (twin SPTs).  \label{fig:threephases}}
\end{figure}
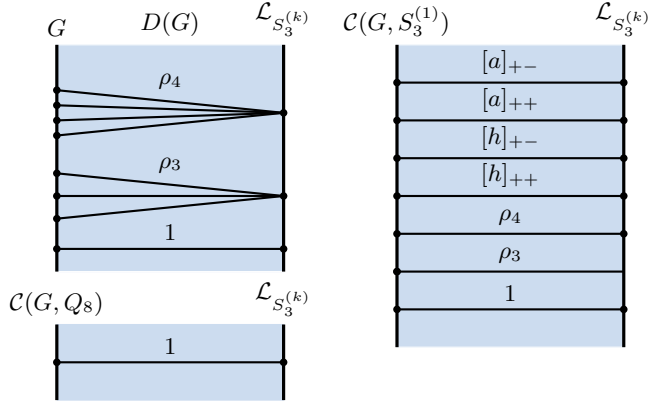

\noindent\textbf{Gauging $Q_8$.}
The symmetry category $\cC(G, Q_8) \cong \SymCatQ$, has thirty symmetry operators. Denote the $\Rep(Q_8)$ symmetry generators by
\begin{equation}
    1, \, 1_{iz}, \, 1_{xz}, \, 1_{iz}1_{xz}, \, m. 
\end{equation}
Here $1_{g}$ is the unique non-trivial one-dimensional representation on $Q_8$ such that $1_{g}(g)=1$, $g\in \{iz,xz\}$.
The invertible symmetry operators $1,1_{iz},1_{xz},1_{iz}1_{xz}$ form $\Z_2^{\langle 1_{iz}\rangle}\times \Z_2^{\langle 1_{xz}\rangle}$ while the non-invertible $m$ satisfies
\begin{equation}
    z\otimes m \cong m \otimes z \cong  m, \quad m\otimes m \cong 1\oplus 1_{iz} \oplus 1_{xz} \oplus 1_{iz}1_{xz},
\end{equation}
for any $z\in \{1,1_{iz},1_{xz},1_{iz}1_{xz}\}$. Denote the generators of $S_3^{(1)}$ by $a$ and $h$, $a^3=h^2=1$, $hah^{-1}=a^2$, they act on $\Rep(Q_8)$ as follows:
\begin{equation}
    \begin{split}
        a:\Rep(Q_8)& \rightarrow\Rep(Q_8)\\
        1_{iz} & \mapsto 1_{iz}1_{xz},\\
        1_{xz} & \mapsto 1_{iz},\\
        1_{iz}1_{xz} & \mapsto 1_{xz}, \\
    \end{split} \qquad \begin{split}
        h:\Rep(Q_8)& \rightarrow\Rep(Q_8)\\
        1_{iz} & \mapsto 1_{iz}1_{xz},\\
        1_{xz}1_{xz} & \mapsto 1_{iz}, \\
        & \\
    \end{split} 
\end{equation}
and trivially on the other simple objects.

Denote the $g\in S_3^{(1)}$ action on $z\in \Rep(Q_8)$ by ${}^gz$, and a symmetry operator in $\cC(G, Q_8)\cong \SymCatQ$ by $z \boxtimes g$. Simple symmetry operators form the set $\{z\boxtimes g\,|\,z\in \text{Irr}(\Rep(Q_8)), g\in S_3^{(1)}\}$, and fuse according to
\begin{equation}
    (z_1\boxtimes g_1) \otimes (z_2\otimes g_2) \cong (z_1\otimes {}^{g_1}z_2) \boxtimes (g_1g_2),
\end{equation}
where $z_1,z_2\in \Rep(Q_8)$ and $g_1,g_2\in S_3^{(1)}$.

\vspace{2mm}
\noindent\textbf{Gauging $S_3^{(1)}$.}
The symmetry category $\SymCatS:=\cC(G,S_3^{(1)})$ has seven simple symmetry operators that we denote by
\begin{equation} \label{eq:obj_cS}
    1,\,{\rho_{1-}},\, \rho_{2},\, c,\, c\rho_{1-},\, c\rho_2,\, [iz]\,,
\end{equation}
where $1$, $\rho_{1-}$ and $\rho_2$ satisfy the fusion rule of $\Rep(S_3^{(1)})$, and the fusion rules can be deduced from the following: 
\begin{equation}
\label{eqn:fusionGaugeS3}
    \begin{split}
        \rho_{1-} \otimes \rho_{1-} &\cong 1\,, \quad \rho_{1-}\otimes \rho_2 \cong \rho_{2}\cr  
        \rho_{1}\otimes c &\cong c\rho_{1-}\,, \quad \rho_{2} \otimes c \cong c\rho_2\,, \quad c\otimes c\cong 1\,, \\
         c\otimes [iz] &\cong [iz]\,, \quad \rho_{1-}\otimes [iz] \cong [iz]\,, \quad \rho_{2}\otimes [iz] \cong 2[iz]\,, \\
         \rho_2\otimes \rho_2& \cong 1 \oplus \rho_{1-}\oplus \rho_2\,,\\
         [iz] \otimes [iz] &\cong 1\oplus \rho_{1-} \oplus 2\rho_2 \oplus c \oplus c\rho_{1-} \oplus 2c\rho_2 \oplus 4 [iz]\,.
    \end{split}
\end{equation}
Fig.~\ref{fig:threephases} shows the SymTFTs for fixing the symmetry boundaries of the SymTFTs to be the Dirichlet boundary, boundaries for $\cC(G, Q_8)$ and $\cC(G,S_3^{(1)})$. 

\section{Lattice Model for Twin Phases and Transitions}
\label{app:Lattice}
In this appendix we discuss the details of the anyon chain model for the gapped twin phases and phase transition. 
We consider a spatial circle, with $L$ edges, each of which has basis states $\{\ket{f}:f\in G\}$. These are acted upon by left and right multiplication operators
\be \label{eq:LgRg}
    L^g\ket{f}=\ket{gf}\,,\qquad   R^g\ket{f}=\ket{fg}\,,
\ee
and $\dim(\rho)^2$ diagonal operators for each irrep $\rho$ of $G$ 
\be 
    Z^{\rho}_{n,m}\ket{g}=
    \rho(g)_{n,m}\ket{g}\,,
\ee
where $\rho(g)$ is the unitary matrix representation of $g$ in the irrep $\rho$, and $n,m$ label its row and column indices.

We define a 1+1d lattice model on a spatial circle of length $L$, using the anyon chain see e.g. 
\cite{Feiguin:2006ydp,
Trebst:2008fibonacci,
Gils:2008collective,
Gils:2009topology,
Pfeifer:2010translation,
Ardonne:2010yanglee,
Gils:2013spin1, 
Aasen:2016dop,
Aasen:2020jwb,
Buican:2017anyonic, 
Sinha:2023hum, 
Bhardwaj:2024kvy,
Wen:2026ncw} and denote basis states by $\ket{g_1,g_2,\cdots,g_L}$ with $g_j\in G$ and $j=1,\cdots,L$.

\textbf{Anomalous $G$ symmetry action.} The symmetry operators act from below
\be 
\begin{tikzpicture}
\draw[thick, ->-] (-2, 0) --(0,0) ;
\draw[thick, ->-] (0,0) -- (2,0) ;
\draw[thick, ->-] (0, 0) --(0,2) ;
\node[above] at (-2,0) {$g_j$} ;
\node[above] at (2,0) {$g_{j+1}$} ;
\node[above] at (0,2) {$g_jg_{j+1}^{-1}$} ;
\draw[ForestGreen, thick, ->-] (-2,-0.4) -- (2, -0.4); 
\node[ForestGreen, below] at (0,-0.4) {$g$};
\end{tikzpicture}
\ee 
Fusing this $g$-line onto the chain requires 
first deforming the figure to 
\be 
\begin{tikzpicture}
\draw[thick, ->-] (-2, 0) --(-0.75,0) ;
\draw[thick, ->-] (-0.75, 0) --(0,0) ;
\draw[thick, ->-] (0,0) -- (0.75,0) ;
\draw[thick, ->-] (0.75,0) -- (2,0) ;
\draw[thick, ->-] (0, 0) --(0,2) ;
\node[above] at (-2,0) {$g_jg$} ;
\node[above] at (2,0) {$g_{j+1}g$} ;
\node[above] at (0,2) {$g_jg_{j+1}^{-1}$} ;
\node[above] at (-0.5, 0) {$g_j$} ; 
\node[above] at (0.5, 0) {$g_{j+1}$} ; 
\draw[ForestGreen,thick, ->-] (-0.75,0) arc[start angle=180,end angle=360,radius=0.75];
\draw[ForestGreen, thick , ->-] (-2,0) -- (-0.75, 0); 
\draw[ForestGreen, thick, ->- ] (0.75,0) -- (2, 0); 
\node[ForestGreen, below] at (0,-0.7) {$g$};
\end{tikzpicture}
\ee
and then applying the 
$F$-move with
\be 
F_{g_j g}^{g_j g_{j+1}^{-1}, g_{j+1}, g} = \omega (g_j g_{j+1}^{-1}, g_{j+1}, g) \,,
\ee
which brings it to the form 
\be
\begin{tikzpicture}
\node at (-2.5,1.5) {$\omega (g_j g_{j+1}^{-1}, g_{j+1}, g) \times $};
\draw[thick, ->-] (0, 0) --(0,2) ;
\node[above] at (-1.8,0) {$g_jg$} ;
\node[above] at (2.5,0) {$g_{j+1}g$} ;
\node[above] at (1.5,0) {$g_{j+1}$} ;
\node[above] at (0,2) {$g_jg_{j+1}^{-1}$} ;
\node[above] at (0.5, 0) {$g_{j+1}g$} ; 
\draw[ForestGreen,thick, ->-] (1,0) arc[start angle=180,end angle=360,radius=0.5];
\draw[thick, ->-] (1, 0) --(2,0) ;
\draw[ForestGreen, thick , ->-] (-2,0) -- (0, 0); 
\draw[ForestGreen, thick , ->-] (0,0) -- (1, 0); 
\draw[ForestGreen, thick , ->-] (2,0) -- (3, 0); 
\node[ForestGreen, below] at (1.5,-0.6) {$g$};
\end{tikzpicture}
\ee
Popping the bubble gives back a fusion tree diagram in the Hilbert space with $g\in G$.  
Therefore the symmetry action by $g\in G$ comprises of right multiplication and the anomalous phase $\omega(g_{j}g_{j+1}^{-1},g_{j+1},g)$ between all neighboring sites:
\be\ba  \label{eq:Ugom}
    U_g\ket{g_1,\cdots,g_L}\!=\!\prod_j \omega(g_{j}g_{j+1}^{-1},g_{j+1},g)\ket{g_1g,\cdots,g_Lg}\,.
\ea\ee

\noindent\textbf{Hamiltonian action.} The $G$-symmetric anyon chain Hamiltonian for the phase labeled by the subgroup $K$ (and trivial 2-cocycles) contains operators $L^{f,\om}_{j+1}$ for $f\in K$ acting as:
\be
\begin{tikzpicture}
\begin{scope}[scale=1.1, shift={(6.75,0)}]
\draw [thick,-stealth](-0.5,-0.5) -- (0.25,-0.5);
\draw [thick](0,-0.5) -- (1.5,-0.5);
\node at (0.25,-1) {$g_{j}$};
\begin{scope}[shift={(2,0)}]
\draw [thick,-stealth](-0.5,-0.5) -- (0.25,-0.5);
\draw [thick](0,-0.5) -- (1.25,-0.5);
\node at (0.25,-1) {$g_{j+1}$};
\end{scope}
\begin{scope}[shift={(0.75,0)},rotate=90]
\draw [thick,-stealth,red!70!green](-0.5,-0.5) -- (0.25,-0.5);
\draw [thick,red!70!green](-0.5,-0.5) -- (0.75,-0.5);
\node[red!70!green] at (2.25,-2.8) {$fg_{j+1}g_{j+2}^{-1}$};
\node[red!70!green] at (1,-1.5) {$f$};
\draw[fill=blue!71!green!58] (-0.5,-0.5) ellipse (0.1 and 0.1);
\end{scope}
\begin{scope}[shift={(3.75,0)}]
\draw [thick,-stealth](-0.5,-0.5) -- (0.5,-0.5);
\draw [thick](0.25,-0.5) -- (1,-0.5);
\node at (0.5,-1) {$g_{j+2}$};
\end{scope}
\begin{scope}[shift={(2.75,0)},rotate=90]
\draw [thick,-stealth,red!70!green](-0.5,-0.5) -- (0.25,-0.5);
\draw [thick,red!70!green](-0.5,-0.5) -- (0.75,-0.5);
\node[red!70!green] at (2.25,1.8) {$g_{j}g_{j+1}^{-1}f^{-1}$};
\node[red!70!green] at (0.25,-1.3) {$g_{j+1}g_{j+2}^{-1}$};
\node[red!70!green] at (0.25,2.1) {$g_{j}g_{j+1}^{-1}$};
\draw[fill=blue!71!green!58] (-0.5,-0.5) node (v1) {} ellipse (0.1 and 0.1);
\end{scope}
\draw [thick,-stealth,red!70!green](3.25,1.25) -- (3.25,1.5);
\draw [thick,-stealth,red!70!green](1.25,1.25) -- (1.25,1.5);
\draw [thick,-stealth,red!70!green](2,0.75) -- (2.25,0.75);
\draw [thick,red!70!green](3.25,0.75) -- (3.25,2);
\draw [thick,red!70!green](1.25,0.75) -- (1.25,2);
\draw [thick,red!70!green](3.25,0.75) -- (1.25,0.75);
\draw[fill=blue!60] (3.25,0.75) ellipse (0.1 and 0.1);
\draw[fill=blue!60] (1.25,0.75) ellipse (0.1 and 0.1);
\end{scope}
\end{tikzpicture}
\ee

\be
\begin{tikzpicture}
\begin{scope}[scale=1.1, shift={(6.75,0)}]
\draw [thick,-stealth](-0.5,-0.5) -- (0.25,-0.5);
\draw [thick](0,-0.5) -- (1.5,-0.5);
\node at (0.25,-1) {$g_{j}$};
\begin{scope}[shift={(2,0)}]
\draw [thick,-stealth](-0.5,-0.5) -- (0.25,-0.5);
\draw [thick](0,-0.5) -- (1.25,-0.5);
\node at (0.25,-1) {$fg_{j+1}$};
\end{scope}
\begin{scope}[shift={(0.75,0)},rotate=90]
\draw [thick,-stealth,red!70!green](-0.5,-0.5) -- (0.25,-0.5);
\draw [thick,red!70!green](-0.5,-0.5) -- (0.75,-0.5);
\node[red!70!green] at (2.25,-2.8) {$fg_{j+1}g_{j+2}^{-1}$};
\node[red!70!green] at (0.5,-1.75) {$f$};
\node[red!70!green] at (0.5,-1.25) {$f$};
\draw[fill=blue!71!green!58] (-0.5,-0.5) ellipse (0.1 and 0.1);
\end{scope}
\begin{scope}[shift={(3.75,0)}]
\draw [thick,-stealth](-0.5,-0.5) -- (0.5,-0.5);
\draw [thick](0.25,-0.5) -- (1,-0.5);
\node at (0.5,-1) {$g_{j+2}$};
\end{scope}
\begin{scope}[shift={(2.75,0)},rotate=90]
\draw [thick,-stealth,red!70!green](-0.5,-0.5) -- (0.25,-0.5);
\draw [thick,red!70!green](-0.5,-0.5) -- (0.75,-0.5);
\node[red!70!green] at (2.25,1.8) {$g_{j}g_{j+1}^{-1}f^{-1}$};
\node[red!70!green] at (0.25,-1.3) {$g_{j+1}g_{j+2}^{-1}$};
\node[red!70!green] at (0.25,2.1) {$g_{j}g_{j+1}^{-1}$};
\draw[fill=blue!71!green!58] (-0.5,-0.5) node (v1) {} ellipse (0.1 and 0.1);
\end{scope}
\draw [thick,-stealth,red!70!green](3.25,1.25) -- (3.25,1.5);
\draw [thick,-stealth,red!70!green](1.25,1.25) -- (1.25,1.5);
\draw [thick,red!70!green](3.25,0.75) -- (3.25,2);
\draw [thick,red!70!green](1.25,0.75) -- (1.25,2);
\draw [thick,-stealth,red!70!green](1.75, 0.7) -- (1.75, 0);
\draw [thick,-stealth,red!70!green](2.75, -0.5) -- (2.75, 0.25);
\draw [thick,red!70!green](3.25,0.75) -- (2.75,0.75) -- (2.75,-0.5);
\draw [thick,red!70!green](1.75,-0.5) -- (1.75,0.75) -- (1.25,0.75);
\draw[fill=blue!60] (3.25,0.75) ellipse (0.1 and 0.1);
\draw[fill=blue!60] (1.25,0.75) ellipse (0.1 and 0.1);
\end{scope}
\end{tikzpicture}
\ee

which, using two $F$-moves evaluates to
\be\ba \label{eq:Lgom}
        L_{j+1}^{f,\om}\ket{g_{j},\,g_{j+1},\,g_{j+2}}=
        &\frac{\omega(g_{j}g_{j+1}^{-1}f^{-1},\;f,\;g_{j+1})}{\omega(f,\;g_{j+1}g_{j+2}^{-1},\;g_{j+2})}\times\\
        &\ket{g_{j},\,fg_{j+1},\,g_{j+2}}\,.
\ea\ee

\noindent\textbf{Hamiltonians for Twin gapped phases.}
We will use the following Hilbert space decomposition for the local Hilbert space of $G=GL(2,3)$:
\be \label{eq:Hbits}
\bC[G]\cong \bC^3_{a} \otimes \bC^2_{h} \otimes \bC^2_{hc} \otimes \bC^2_{iz} \otimes \bC^2_{xz}  
\ee
where the subscript labels the corresponding group element. Note that the symmetry operators \eqref{eq:Ugom} for $h,hc$, restricted to their two qubits, simplify to:
\be\ba
    U_h|_{\Z_2^{\langle h\rangle}\times\Z_2^{\langle hc\rangle}}&=\prod_jX^h_j\,,\\
    U_{hc}|_{\Z_2^{\langle h\rangle}\times\Z_2^{\langle hc\rangle}}&=\prod_jX^{hc}_j(-1)^{(\bbI+Z^h)_j(\bbI-Z^h)_{j+1}/4}
\ea\ee
from which we see that $U_{hc}$ carries an $h$-charge, acting with $-1$ on states with an $h$-domain wall of the type $g_j|_{\bC^2_h}=\id$ and $g_{j+1}|_{\bC^2_h}=h$.

Using the explicit representation matrices provided in App.~\ref{app:notations} and denoting  
\be \label{eq:PS3k}
    \left(Z^{\rho}_{j}\cdot Z^{\rho\dagger}_{j+1}\right)_{n,n}=\sum_{m=1}^{\dim(\rho)}(Z^\rho_{n,m})_j (\ol{Z^\rho_{n,m}})_{j+1}\,,
\ee
the local operator enforcing spontaneous symmetry breaking of $G\to S_3^{(k)}$, i.e. projecting onto the subspace of states for which $g_{j}g_{j+1}^{-1}\in S_3^{(k)}$, is:
\be\ba \label{eq:projS3k}
    P^{(S_3^{(k)})}_{j+\half}\!\!=\!\frac{1}{8}\!\left[\mathbb I_{j,j+1}\!
    +3\!\left(Z^{\rho_3}_{j}\cdot Z^{\rho_3\dagger}_{j+1}\right)_{1,1}\!\!\!\!
    +4\!\left(Z^{\rho_4}_{j}\cdot Z^{\rho_4\dagger}_{j+1}\right)_{k,k}\right]. 
\ea\ee
The entry of $\rho_4$ is $(1,1)$ and $(2,2)$ respectively and corresponds to the trivial $S_3^{(k)}$ irrep in the decomposition
\be 
    \rho_4|_{S_3^{(1)}}= 1\oplus \rho_{1-} \oplus \rho_2\,,\quad 
    \rho_4|_{S_3^{(2)}}= \rho_{1-} \oplus 1 \oplus \rho_2\,.
\ee

The disordering operator enforcing that $S_3^{(k)}$ is preserved for $k=1,2$ is, for site $j$:
\be\ba
    L^{(S_3^{(k)}\!\!,\om)}_{j}&=\frac{1}{6}\lb\bbI_{j}+L_{j}^{a,\om}+L_{j}^{a^2,\om}\rb\lb\bbI_{j}+L_{j}^{hc^{k-1},\om}\rb\\
    &\simeq \frac{1}{3}\lb\bbI_{j}+L_{j}^{a,\om}+L_{j}^{a^2,\om}\rb+\frac{1}{2}\lb\bbI_{j}+L_{j}^{hc^{k-1}\!,\om}\rb
\ea\ee
where the individual multiplication operators were defined in \eqref{eq:Lgom} (for site $j+1$) and we wrote a simplified form in the second line. Note that, for group elements in $D_{12}=\langle S_3^{(1)}, S_3^{(2)}\rangle$ our 3-cocycle is
\be \label{eq:omD12}
\om|_{D_{12}}(g_1,g_2,g_3)=\om_{\II}(p(g_1),p(g_2),p(g_3))
\ee
where $p:D_{12}\to D_{12}/\Z_3=\langle p(h),p(hc)\rangle\cong\Z_2\times\Z_2$. \\[1mm]

The Hamiltonians for the twin gapped phases labeled by $S_3^{(k)}$ for $k=1,2$ are therefore:
\be\ba \label{eq:HG_gapped}
    H_{S_3^{(k)}}=&-\sum_j\Big(P^{(S_3^{(k)})}_{j-\half}
    L^{(S_3^{(k)}\!\!,\om)}_{j}
    P^{(S_3^{(k)})}_{j+\half}\Big)  +\text{h.c.}\\
    \simeq&-\sum_j\Big(L^{(S_3^{(k)}\!\!,\om)}_{j}+
    P^{(S_3^{(k)})}_{j+\half}\Big) +\text{h.c.} \\
    \simeq&-\sum_j\frac{1}{8}\left[\mathbb I_{j,j+1}\!
    +3\!\left(Z^{\rho_3}_{j}\cdot Z^{\rho_3\dagger}_{j+1}\right)_{1,1}\right]_j\\
    &-\sum_j\frac{1}{3}\lb\bbI_j+L_j^{a,\om}+L_j^{a^2,\om}\rb-\sum_j\frac{1}{2}\lb\bbI+L_j^{hc^{k-1}\!,\om}\rb \\
    &-\sum_j\frac{1}{2}\left[\!\left(Z^{\rho_4}_{j}\cdot Z^{\rho_4\dagger}_{j+1}\right)_{k,k}\right]_j +\text{h.c.}
\ea\ee
(where we add the hermitian conjugate h.c. to obtain Hermitian Hamiltonians).

The ground state corresponding to the trivial coset is obtained by acting with the disordering operators on the identity state:
\be\ba
   \GSrz=&\prod_jL^{(S_3^{(k)}\!\!,\om)}_{j}\bigotimes_j\ket{\id}_j=\bigotimes_j\frac{1}{\sqrt{6}}\sum_{g\in S_3^{(k)}}\ket{g}_j
\ea\ee
where we used the property that $\om|_{S_3^{k)}}\equiv1$ so the operators $L_j^{(g,\om)}$ for $g\in S_3^{(k)}$ simplify to $L_j^g$ when acting on states with all group elements in $S_3^{(k)}$. Explicitly, using the Hilbert space decomposition \eqref{eq:Hbits}, they are:
\be\ba
    |\GS^{(1)},\id\rangle&=\bigotimes_j(\ket{+}\ket{+}\ket{0}\ket{0}\ket{0})_j\,,\\[-2mm]
    |\GS^{(2)},\id\rangle&=\bigotimes_j(\ket{+}\ket{0}\ket{+}\ket{0}\ket{0})_j \\[-3mm]
\ea\ee
where, for the qutrit $\ket{+}:=\frac{1}{\sqrt{3}}(\ket{\id}+\ket{a}+\ket{a^2})$ and for qubits, $\ket{+}=\frac{1}{\sqrt{2}}(\ket{\id}+\ket{g})$ with $g=h,hc$ the $\Z_2$ generator. $\GSrz$ is clearly invariant under $U_g$ for $g\in S_3^{(k)}$. The other ground states are obtained by acting on $\GSrz$ with the SSB'ed generators
\be
   \GSr=U_{q}\GSrz\,, \quad q\in Q_8\,.
\ee
and preserve the symmetries $U_{q^{-1}gq}$ for $g\in S_3^{(k)}$.

The operators \eqref{eq:PS3k} provide the local OPs with $+1$ vev in the ground states for phase $k=1,2$:
\be\ba
    \GSkl \left(Z^{\rho_3}_{j}\cdot Z^{\rho_3\dagger}_{j+1}\right)_{1,1} \GSkr=1\,,\\
    \GSkl \left(Z^{\rho_4}_{j}\cdot Z^{\rho_4\dagger}_{j+1}\right)_{k,k} \GSkr=1\,,
\ea\ee
where $\GSkr$ is any state in the ground state subspace of $H_{S_3^{(k)}}$. The second line shows that the operator gaining a vev is a different component of $\rho_4$ for the two twin phases $k=1,2$. 

\vspace*{1mm}
\noindent\textbf{Phase transition Hamiltonian.} The twin phases are distinct gapped phases (since they preserve non-conjugate symmetries). To describe the transiton between them, we consider the linear interpolation between the Hamiltonians \eqref{eq:HG_gapped} with $\lambda\in [0,1]$:
\be\ba
    H(\lambda)=&(1-\lambda)H_{S_3^{(1)}}+\lambda H_{S_3^{(2)}}\\
    \simeq&-\sum_j\frac{1}{8}\left[\mathbb I_{j,j+1}\!
    +3\!\left(Z^{\rho_3}_{j}\cdot Z^{\rho_3\dagger}_{j+1}\right)_{1,1}\right]_j \\
    &-\frac{1}{3}\sum_j\lb\bbI_j+L_j^{a,\om}+L_j^{a^2,\om}\rb -\frac{1}{2}\sum_j\bbI_j  \\
    &-\frac{1}{2}\sum_j (1-\lambda)\lbb L_j^{h,\om}+\left(Z^{\rho_4}_{j}\cdot Z^{\rho_4\dagger}_{j+1}\right)_{1,1}\rbb \\
    &-\frac{1}{2}\sum_j \lambda\lbb L_j^{hc,\om}+\left(Z^{\rho_4}_{j}\cdot Z^{\rho_4\dagger}_{j+1}\right)_{2,2}\rbb  +\text{h.c.}
\ea\ee
The system will undergo a phase transition, with the critical point occurring at $\lambda=1/2$. 

Recalling the Hilbert space decomposition \eqref{eq:Hbits}, the terms in the Hamiltonian act as follows: the first line preserves $D_{12}$ and spontaneously breaks the $Q_8$ generators $iz,xz$, the second line projects the qutrit onto its $\ket{+}$ state, while the last two lines act non-trivially on the qubits for $h$ and $hc$, which carry the mixed anomaly $\om$. 

\clearpage
\onecolumngrid

\section{Hasse Diagram of Condensable Algebras of $\cZ^\omega(GL(2,3))$}\label{app:Hasse}

\begin{table}[H]
\renewcommand{\arraystretch}{1.2}
\begin{center}
\begin{tabular}{|c|c|c|c|}\hline
$d$ & Name & \hfill$A(H,N,\gamma,\epsilon)$ and Anyon Decomposition\hfill\; & Reduced TO \\ \hline

1 &
\nameref{alg:1}\xlabel[$A^\om_{1}$]{alg:1} &
$A(GL(2,3),1,1,1)\cong_{\obj}1$ &
$D({GL(2,3)^\omega})$
\\ \hline

2 &
\nameref{alg:2}\xlabel[$A^\om_{2}$]{alg:2} &
$A(SL(2,3),1,1,1)\cong_{\obj} 1 \oplus \rho_1$ &
$D({SL(2,3)^\alpha})$
\\ \hline

3 &
\nameref{alg:3}\xlabel[$A^\om_{3}$]{alg:3} &
$A(QD_{16},1,1,1)\cong_{\obj} 1 \oplus \rho_2$ &
$D({QD_{16}^\alpha})$
\\ \hline

4 &
\nameref{alg:4}\xlabel[$A^\om_{4}$]{alg:4} &
$A(D_{12},1,1,1)\cong_{\obj} 1 \oplus \rho_3$ &
$D({D_{12}^\alpha})$
\\ \hline

6 &
\nameref{alg:5}\xlabel[$A^\om_{5}$]{alg:5} &
$A(\Z_8,1,1,1)\cong_{\obj}1 \oplus \rho_2 \oplus \rho_{3-} $ &
$D({\Z_8^\alpha})$
\\ \hline

6 &
\nameref{alg:6}\xlabel[$A^\om_{6}$]{alg:6} &
$A(D_8,1,1,1)\cong_{\obj} 1 \oplus \rho_2 \oplus \rho_3$ &
$D({D_8^\alpha})$
\\ \hline

6 &
\nameref{alg:7}\xlabel[$A^\om_{7}$]{alg:7} &
$A(Q_8,1,1,1)\cong_{\obj} 1 \oplus \rho_1 \oplus 2\rho_2$ &
$D({Q_8^\alpha})$
\\ \hline

8 &
\nameref{alg:8}\xlabel[$A^\om_{8}$]{alg:8} &
$A(\Z_6,1,1,1)\cong_{\obj}1 \oplus \rho_1 \oplus \rho_{3-} \oplus \rho_3 $ &
$D({\Z_6^\alpha})$
\\ \hline

8 &
\nameref{alg:9}\xlabel[$A^\om_{9}$]{alg:9} &
$A(S_3^{(1)},1,1,1)\cong_{\obj}1 \oplus \rho_3 \oplus \rho_4 $ &
$D({S_3})$
\\ \hline

8 &
\nameref{alg:10}\xlabel[$A^\om_{10}$]{alg:10} &
$A(S_3^{(2)},1,1,1)\cong_{\obj}1 \oplus \rho_3 \oplus \rho_4 $ &
$D({S_3})$
\\ \hline

12 &
\nameref{alg:11}\xlabel[$A^\om_{11}$]{alg:11} &
$A(D_{12},\Z_3,1,1)\cong_{\obj} 1 \oplus \rho_3 \oplus [a]_{++}$ &
$D({(\Z_2\times \Z_2)}^{\om_\II})$
\\ \hline

12 &
\nameref{alg:12}\xlabel[$A^\om_{12}$]{alg:12} &
$A(\Z_2\times \Z_2,1,1,1)\cong_{\obj}1 \oplus \rho_2 \oplus \rho_{3-} \oplus 2\rho_3 $ &
$D({(\Z_2\times \Z_2)}^{\om_\II})$
\\ \hline

12 &
\nameref{alg:13}\xlabel[$A^\om_{13}$]{alg:13} &
$A(\Z_4,1,1,1)\cong_{\obj} 1 \oplus \rho_1 \oplus 2\rho_2 \oplus \rho_{3-} \oplus \rho_3$ &
$D({\Z_4^\alpha})$
\\ \hline

16 &
\nameref{alg:14}\xlabel[$A^\om_{14}$]{alg:14} &
$A(\Z_3,1,1,1)\cong_{\obj} 1 \oplus \rho_1 \oplus \rho_{3-} \oplus \rho_3 \oplus 2\rho_4$ &
$D({\Z_3})$
\\ \hline

24 &
\nameref{alg:15}\xlabel[$A^\om_{15}$]{alg:15} &
$A(\Z_6,\Z_3,1,1)\cong_{\obj} 1 \oplus \rho_1 \oplus \rho_{3-} \oplus \rho_3 \oplus 2[a]_{++}$ &
$D({\Z_2^\alpha})$
\\ \hline

24 &
\nameref{alg:16}\xlabel[$A^\om_{16}$]{alg:16} &
$A(S_3^{(1)},\Z_3,1,1)\cong_{\obj} 1 \oplus \rho_3 \oplus \rho_4 \oplus [a]_{++} \oplus [a]_{+-}$ &
$D({\Z_2})$
\\ \hline

24 &
\nameref{alg:17}\xlabel[$A^\om_{17}$]{alg:17} &
$A(S_3^{(2)},\Z_3,1,1)\cong_{\obj} 1 \oplus \rho_3 \oplus \rho_4 \oplus [a]_{++} \oplus [a]_{+-}$ &
$D({\Z_2})$
\\ \hline

24 &
\nameref{alg:18}\xlabel[$A^\om_{18}$]{alg:18} &
$A(\Z_2,1,1,1)\cong_{\obj} 1 \oplus \rho_2 \oplus \rho_{2+} \oplus \rho_{2-} \oplus \rho_{3-} \oplus 2\rho_3 \oplus 2\rho_4$ &
$D({\Z_2})$
\\ \hline

24 &
\nameref{alg:19}\xlabel[$A^\om_{19}$]{alg:19} &
$A(\Z_2,1,1,1)\cong_{\obj} 1 \oplus \rho_1 \oplus 2\rho_2 \oplus 3\rho_{3-} \oplus 3\rho_3$ &
$D({\Z_2^\alpha})$
\\ \hline\hline

48 &
\nameref{alg:20}\xlabel[$A^\om_{20}$]{alg:20} &
$\textcolor{oxfordblue}{A(S_3^{(1)},S_3^{(1)},1,1)}\cong_{\obj} 1 \oplus \rho_3 \oplus \rho_4 \oplus [h]_{+-} \oplus [h]_{++} \oplus [a]_{++} \oplus [a]_{+-}$ &
Trivial
\\ \hline

48 &
\nameref{alg:21}\xlabel[$A^\om_{21}$]{alg:21} &
$\textcolor{oxfordblue}{A(S_3^{(2)},S_3^{(2)},1,1)}\cong_{\obj} 1 \oplus \rho_3 \oplus \rho_4 \oplus [h]_{+-} \oplus [h]_{++} \oplus [a]_{++} \oplus [a]_{+-}$ &
Trivial
\\ \hline

48 &
\nameref{alg:22}\xlabel[$A^\om_{22}$]{alg:22} &
$A(\Z_3,\Z_3,1,1)\cong_{\obj} 1 \oplus \rho_1 \oplus \rho_{3-} \oplus \rho_3 \oplus 2\rho_4 \oplus 2[a]_{++} \oplus 2[a]_{+-}
$ &
Trivial
\\ \hline

48 &
\nameref{alg:23}\xlabel[$A^\om_{23}$]{alg:23} &
$A(\Z_2,\Z_2,1,1)\cong_{\obj} 1 \oplus \rho_2 \oplus \rho_{2+} \oplus \rho_{2-} \oplus \rho_{3-} \oplus 2\rho_3 \oplus 2\rho_4 \oplus [h]_{+-} \oplus [h]_{++}$ &
Trivial
\\ \hline

48 &
\nameref{alg:24}\xlabel[$A^\om_{24}$]{alg:24} &
$A(1,1,1,1)\cong_{\obj} 1 \oplus \rho_1 \oplus 2\rho_2 \oplus 2\rho_{2+} \oplus 2\rho_{2-} \oplus 3\rho_{3-} \oplus 3\rho_3 \oplus 4\rho_4
$ &
Trivial
\\ \hline

\end{tabular}
\end{center}
\caption{Condensable algebras for $D^\omega(GL(2,3))$, with their dimension, label, algebra data, anyon decomposition and reduced TO (here, $\alpha$ is a non-trivial 3-cocycle). The twin algebras are Lagrangian algebra twins $A^\om_{20}$ and $A^\om_{21}$ and non-maximal twins $A^\om_{16}, A^\om_{17}$ and $A^\om_{9}, A^\om_{10}$.}
\label{tab:Galg-short}
\end{table}

\onecolumngrid

\clearpage
\newpage

\begin{figure}[H]
\begin{center}
\begin{tikzpicture}[vertex/.style={draw}, scale=0.9]
 \begin{scope}[shift={(0,0)}]
    \foreach \coord/\i/\j in {
(0,1)/1/{\nameref{alg:1}},
(4.,-1)/2/{\nameref{alg:2}},
(1,-2)/3/{\nameref{alg:3}},
(-2.5,-3)/4/{\nameref{alg:4}},
(2.5,-4)/5/{\nameref{alg:5}},
(1,-4)/6/{\nameref{alg:6}},
(4,-4)/7/{\nameref{alg:7}},
(6,-6.5)/8/{\nameref{alg:8}},
(-3.5,-6)/9/{\nameref{alg:9}},
(-1.5,-6)/10/{\nameref{alg:10}},
(-6,-8)/11/{$\cA(H, N)=$\nameref{alg:11}},
(1.5,-8)/12/{\nameref{alg:12}},
(5,-8)/13/{\nameref{alg:13}},
(1,-11)/14/{\nameref{alg:14}},
(-3,-13)/15/{\nameref{alg:15}},
(-6,-13)/16/{\nameref{alg:16}},
(1,-13)/17/{\nameref{alg:17}},
(4,-13)/18/{\nameref{alg:18}},
(6,-13)/19/{\nameref{alg:19}},
(-6,-16)/20/{$\cL_{S_3^{(1)}}=$\nameref{alg:20}},
(1,-16)/21/{$\cL_{S_3^{(2)}}=$\nameref{alg:21}},
(-3,-16)/22/{\nameref{alg:22}},
(4,-16)/23/{\nameref{alg:23}},
(6,-16)/24/{\nameref{alg:24}}}
    {
      \node[vertex,align=center] (p\i) at \coord {\j};
     }
  \foreach [count=\r] \row in 
{{0,1,1,1,0,0,0,0,0,0,0,0,0,0,0,0,0,0,0,0,0,0,0,0},
{0,0,0,0,0,0,1,1,0,0,0,0,0,0,0,0,0,0,0,0,0,0,0,0},
{0,0,0,0,1,1,1,0,0,0,0,0,0,0,0,0,0,0,0,0,0,0,0,0},
{0,0,0,0,0,0,0,1,1,1,1,1,0,0,0,0,0,0,0,0,0,0,0,0},
{0,0,0,0,0,0,0,0,0,0,0,0,1,0,0,0,0,0,0,0,0,0,0,0},
{0,0,0,0,0,0,0,0,0,0,0,1,1,0,0,0,0,0,0,0,0,0,0,0},
{0,0,0,0,0,0,0,0,0,0,0,0,1,0,0,0,0,0,0,0,0,0,0,0},
{0,0,0,0,0,0,0,0,0,0,0,0,0,1,1,0,0,0,1,0,0,0,0,0},
{0,0,0,0,0,0,0,0,0,0,0,0,0,1,0,1,0,1,0,0,0,0,0,0},
{0,0,0,0,0,0,0,0,0,0,0,0,0,1,0,0,1,1,0,0,0,0,0,0},
{0,0,0,0,0,0,0,0,0,0,0,0,0,0,1,1,1,0,0,0,0,0,0,0},
{0,0,0,0,0,0,0,0,0,0,0,0,0,0,0,0,0,1,1,0,0,0,0,0},
{0,0,0,0,0,0,0,0,0,0,0,0,0,0,0,0,0,0,1,0,0,0,0,0},
{0,0,0,0,0,0,0,0,0,0,0,0,0,0,0,0,0,0,0,0,0,1,0,1},
{0,0,0,0,0,0,0,0,0,0,0,0,0,0,0,0,0,0,0,0,0,1,0,0},
{0,0,0,0,0,0,0,0,0,0,0,0,0,0,0,0,0,0,0,1,0,1,0,0},
{0,0,0,0,0,0,0,0,0,0,0,0,0,0,0,0,0,0,0,0,1,1,0,0},
{0,0,0,0,0,0,0,0,0,0,0,0,0,0,0,0,0,0,0,0,0,0,1,1},
{0,0,0,0,0,0,0,0,0,0,0,0,0,0,0,0,0,0,0,0,0,0,0,1},
{0,0,0,0,0,0,0,0,0,0,0,0,0,0,0,0,0,0,0,0,0,0,0,0},
{0,0,0,0,0,0,0,0,0,0,0,0,0,0,0,0,0,0,0,0,0,0,0,0},
{0,0,0,0,0,0,0,0,0,0,0,0,0,0,0,0,0,0,0,0,0,0,0,0},
{0,0,0,0,0,0,0,0,0,0,0,0,0,0,0,0,0,0,0,0,0,0,0,0},
{0,0,0,0,0,0,0,0,0,0,0,0,0,0,0,0,0,0,0,0,0,0,0,0}}
    {
     \foreach [count=\c] \cell in \row{
            \ifnum\cell=1%
                \draw[-stealth] (p\r) edge [thick] (p\c);
            \fi
        }
    }
\end{scope}
\end{tikzpicture}
\caption{Hasse diagram for $D^\om(G)$. The algebras are ordered by increasing quantum dimension from top to bottom, and are detailed in Tab.~\ref{tab:Galg-short}. The top most algebra is the identity algebra. The bottom row are the Lagrangian algebras, that correspond to topological boundary conditions. $A^\om_{20}$ and $A^\om_{21}$ are the twin Lagrangian algebras. The details of the algebras are in table \ref{tab:Galg-short}.
}
\label{fig:HasseG}
\end{center}
\end{figure}
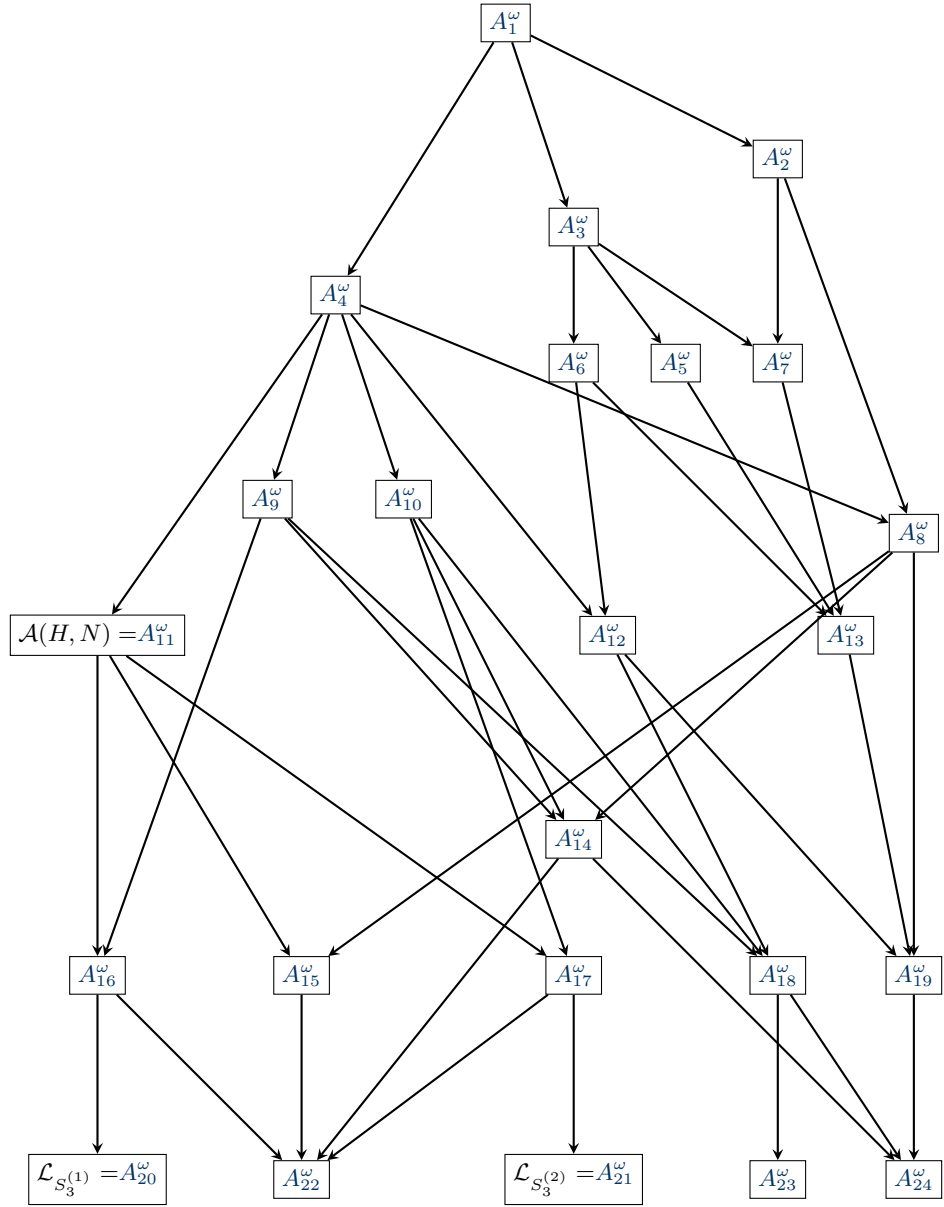

\end{document}